\documentclass[a4paper,fleqn,usenatbib]{mnras}

\usepackage{mathptmx}
\usepackage[T1]{fontenc}
\usepackage{ae,aecompl}

\usepackage{graphicx}	% Including figure files
\usepackage{amsmath}	% Advanced maths commands
\usepackage{amssymb}	% Extra maths symbols

\newcommand{\feii}{Fe\,{\sc ii}}
\newcommand{\phl}{PHL\,1092}
\newcommand{\caii}{Ca\,{\sc ii}}
\newcommand{\oi}{O\,{\sc i}}
\newcommand{\civ}{C\,{\sc iv}}
\newcommand{\oiii}{[O\,{\sc iii}]}
\newcommand{\siii}{[S\,{\sc iii}]}
\newcommand{\ha}{H$\alpha$}
\newcommand{\hb}{H$\beta$}
\newcommand{\nii}{[N\,{\sc ii}]}

\newcommand{\pab}{Pa$\beta$}
\newcommand{\pag}{Pa$\gamma$}
\newcommand{\lya}{Ly$\alpha$}
\newcommand{\hei}{He\,{\sc i}}

\newcommand{\msol}{M$_{\odot}$}
\newcommand{\rop}{R$_{4570}$}
\newcommand{\rni}{R$_{9200}$}
\newcommand{\ron}{R$_{1\mu m}$}
\newcommand{\izwi}{I\,Zw\,1}

\title[Extreme Strong \feii\ Emitter \phl]
{Panchromatic Properties of the Extreme \feii\ Emitter \phl}

\author[Murilo Marinello et al.]
{Murilo Marinello$^{1}$\thanks{E-mail: murilo.marinello@gmail.com},
Alberto Rodr\'iguez-Ardila$^{1,2}$,
Paola Marziani$^{3}$,
Aaron Sigut$^{4}$,\newauthor
Anil Pradhan$^{5}$
\\
$^{1}$Laborat\'{o}rio Nacional de Astrof\'{i}sica -- Rua dos Estados Unidos 154, Bairro das Na\c{c}\~oes. CEP 37504-364, Itajub\'{a}, MG, Brazil\\
$^{2}$Divis\~ao de Astrof\'{\i}sica - INPE, Avenida dos Astronautas 1758, 12227-010, S.J.Campos-SP, Brazil\\
$^{3}$INAF, Osservatorio Astronomico di Padova, vicolo dell Osservatorio 5, IT 35122, Padova, Italy\\
$^{4}$Department of Physics and Astronomy, The University of Western Ontario, London, Ontario N6A 3K7, Canada\\
$^{5}$McPherson Laboratory, The Ohio State University, 140 W. 18th Ave., Columbus, OH 43210-1173, USA\\
 }

\date{Accepted XXX. Received YYY; in original form ZZZ}

\pubyear{2018}

\begin{document}
\label{firstpage}
\pagerange{\pageref{firstpage}--\pageref{lastpage}}
\maketitle

\begin{abstract}
We present near-infrared spectroscopy of the NLS1 galaxy \phl\ ($z=0.394$), the strongest \feii\ emitter ever reported, combined with optical and UV data. We modeled the continuum and the broad emission lines using a power-law plus a black body function and Lorentzian functions, respectively. The strength of the \feii\ emission was estimated using the latest \feii\ templates in the literature. We re-estimate the ratio between the \feii\ complex centered at 4570\AA{} and the broad component of H$\beta$, \rop, obtaining a value of 2.58, nearly half of that previously reported (\rop=6.2), but still placing \phl\ among extreme \feii\ emitters. The FWHM found for low ionization lines are very similar (FWHM$\sim$1200\,km\,s$^{-1}$), but significantly narrower than those of the Hydrogen lines (FWHM$_{\rm H\beta}\sim$1900\,km\,s$^{-1}$). Our results suggest that the \feii\ emission in \phl\ follows the same trend as in normal \feii\ emitters, with \feii\ being formed in the outer portion of the BLR and co-spatial with \caii, and \oi, while \hb\ is formed closer to the central source. The flux ratio between the UV lines suggest high densities, log(n$_{\rm H}$)$\sim13.0$\,cm$^{-3}$, and a low ionization parameter, log(U)$\sim-3.5$. The flux excess found in the \feii\ bump at 9200\AA{} after the subtraction of the NIR \feii\ template and its comparison with optical \feii\ emission suggests that the above physical conditions optimize the efficiency of the \lya-fluorescence process, which was found to be the main excitation mechanism in the \feii\ production. We discuss the role of \phl\ in the Eigenvector 1 context.
\end{abstract}


%&&&&&&&&&&&&&&&&&&&&&&&&&&&&&&&&&&&&&&&&&&&&&&&&%
%&&&&&&&&&&&&&&&% BODY OF PAPER %&&&&&&&&&&&&&&&&%
%&&&&&&&&&&&&&&&&&&&&&&&&&&&&&&&&&&&&&&&&&&&&&&&&%

%%%%%%%%%%%%%%%%%%%%%%%%%%%%%%%%%%%%%%%%%%%%%%%%%%
%%%%%%%%%%%%%%%%% INTRODUCTION %%%%%%%%%%%%%%%%%%%
%%%%%%%%%%%%%%%%%%%%%%%%%%%%%%%%%%%%%%%%%%%%%%%%%%
\section{Introduction}

Narrow-line Seyfert 1 (NLS1) galaxies are a particular subclass of active galactic nuclei (AGN) that show narrow permitted emission lines ($\rm FWHM_{H\beta} < 2000\,km/s$) and weak \oiii$\lambda\lambda$4959,5007 lines (\oiii/\hb~$<3$) \citep{osterbrock85,goodrich89} than classical Type~I AGN.
Besides, NLS1 have other interesting properties across the electromagnetic spectrum. 
In the optical and UV they show strong asymmetries in the \oiii\ lines \citep{zamanov02,bian05,boroson05} and high velocity blueshifted \civ\ lines \citep{sulentic00,sulentic02,wills00,leighly04}.
One of the most intriguing properties of NLS1 is the strong \feii\ emission, with numerous multiplets across the UV, optical and NIR spectrum that form a pseudo-continuum mostly in the UV and optical regions. 
The strength of this feature, usually represented by the flux ratio \ion{Fe}{ii}/H$\beta$ between  the \feii\ bump centred at $\lambda$4570~\AA\ and the broad component of \hb\ (hereafter \rop), is about twice or larger than that measured in normal AGN \citep{zhou06}, with NLS1 reaching values higher than 1 \citep{joly91,marziani01,shen11,rakshit17}.  
  
Several models have tried to explain the \feii\ emission in AGN. 
The most recent ones incorporate excitation mechanisms such as collisional excitation, continuum fluorescence, self-fluorescence of \feii\ lines and Ly$\alpha$-fluorescence \citep{sigut98,verner99,sigut03,bruhweiler08,pradhan11}. 
A decade ago, \citet{bruhweiler08} considerably improved this approach by considering an \feii\ ion with 830 levels, noticing that an appropriate number of energy levels must be taken into account in order to reproduce the observed \feii\ emission.
Understanding this emission in AGN is important at least for 4 reasons: 
($i$) the \feii\ emission is one of the strongest cooling agents in the broad line region (BLR). 
It accounts for up to 25$\%$ of the total energy output of that emission region \citep{wills85}; 
($ii$) it represents a strong component all over the spectrum. 
The myriads of \feii\ multiplets forms a pseudo-continuum from the UV to NIR \citep{sigut03} with more than 344,000 transitions \citep{bruhweiler08}.
($iii$) The gas emitting \feii\ probes the structure and kinematics of the BLR \citep{kovacevic10,hu08,sameshima11,aldama15,cracco16}.  
($iv$) The strength of \feii\ relative to the peak of \oiii\ and width of the \hb\ line forms the optical plane of the Eigenvector 1 (hereafter E1) \citep{boroson92}. 
It is believed to be associated to fundamental properties of AGN such as the accretion power and super massive black hole mass \citep{boroson92,sulentic00,shen14}. 

\citet{boroson92} studied correlations among observed features in a complete radio-quiet sample of quasars using the Principal component analysis (PCA) technique. 
They found that most of the quasar properties are related to Eigenvector 1. 
E1 was later expanded to a 4-dimension eigenvector 1 (4DE1) parameter system to include the blueshift of \civ\ and the X-ray photon index \citep{sulentic00,sulentic02,sulentic07}.
The 4DE1 defines a main sequence for quasars that is believed to be ultimately driven by the Eddington ratio (L/LEdd) \citep{marziani01,shen14} and that, in addition to the four parameters of 4DE1, is correlated with low- and high-ionisation line profiles, systematic shifts of broad and narrow high-ionization lines \citep[see][for a review]{marziani18}.
Moreover, it is possible to identify two populations of type-I AGN in the optical plane of the 4DE1 (defined by the FWHM(\hb) and the \rop): ``Population A" AGN, with FWHM(\hb)$<4000$\,km\,s$^{-1}$ and strong \feii\ emission; and ``Population B" AGN, characterized by FHWM(\hb)$>4000$\,km\,s$^{-1}$ and weaker \feii\ emission \citep{sulentic00a,sulentic00b,sulentic02}.
More recently, \citet{panda18} showed that the Eddington ratio is not enough to drive the quasars through the E1 sequence, suggesting that higher abundances and high densities may also be necessary to place AGN at the high \rop\ end of the E1.
\citet{marziani01}, while looking for a physical interpretation of the E1, pointed out to one particular AGN --\phl--  as an extreme outlier in the high-end of strong \feii\ emitters. 
They considered this source as a rare case that should probe extreme conditions in the BLR.
 
\phl\ is a relatively nearby radio-quiet quasar ($z=0.396$) with many interesting properties all over the electromagnetic spectrum.	
Its X-ray emission is known as one of the weakest and variable in a non-BAL AGN spectrum \citep{miniutti09}.
In the UV, broad \civ\ and \lya\ lines, with a high blueshifted and asymmetric profile in \civ\ are observed \citep{miniutti12}.
The Near-UV spectrum is dominated by the pseudo-continuum formed by \feii.
In the optical region its outstanding \feii\ emission reaches extreme values, with \rop\ varying from 1.81 to 6.2 \citep{lawrence97,bergeron80}. The main cause of this extreme variation is attributed to the method employed in measuring the optical \feii\ emission.
The narrowness of the Balmer lines (FWHM$\sim$1800km/s) and the low ratio \oiii/\hb$\sim$0.9, together with the strong \feii\ emission, classify this AGN as a high luminosity NLS1 \citep{osterbrock85}.
Although \phl\ was already observed  in many regions of the electromagnetic spectrum, to the best of our knowledge, no report of its properties in the NIR exists up to today. 

The NIR has several advantages over the UV and optical for the study of the \feii\ emission. For instance, the most conspicuous \feii\ lines are nearly isolated features, allowing an accurate analysis of their intensities and line profiles.
Moreover, emission lines, such \oi\ and \caii\   are isolated or moderately blended with adjacent lines \citep{aldama15,marinello16}.
While in the optical \hb\ is severely affected by the underlying \feii, the Paschen lines in the NIR are completely isolated \citep{ardila02}.
As pointed out by \citet{sigut98,sigut03,ardila02} and \citep{,marinello16}, this spectral window holds key features for understanding the physics of \feii\ in AGN.
For example, the \feii\ bump centered at $\lambda$9200 and the four isolated \feii\ lines at
$\lambda$9998, 10502, 10900, 11127~\AA\ (known collectively as the 1~$\mu$m \feii\ lines), carry important information about the excitation mechanisms of this ion \citep{sigut98,sigut03,marinello16}.
Therefore optical and NIR observations combined improves the description of the excitation mechanism behind the extreme \feii\ emission in this source.
 
In this paper we analyze for the first time NIR spectroscopy of \phl\ in combination with optical observations around the \hb\ region in order to obtain a full coverage of the most important \feii\ features of this source.
Our goals are threefold: 
($i$) To obtain a consistent estimation of the \feii\ intensity; 
($ii$) to measure the \feii\ and other BLR features (\ion{H}{i}, \ion{O}{i} and \ion{Ca}{ii}) in order to get clues about the location, excitation mechanism, and physical conditions of the regions where they are emitted;
($iii$) To study the role of the \phl\ in the E1 context.
 
This paper is structured as follow: 
Section~2 presents the observations and data reduction. 
Section~3 describes the technique applied to fit the \feii\ templates and measuring the emission lines properties of the broad lines in the observed spectra. 
Section~3 and 4 discusses the location of the \feii\ emitting region, its physical conditions and the excitation mechanism driving this emission.
The role of \phl\ in the E1 context is presented in Section~5.
Conclusion are given in section~6.

%%%%%%%%%%%%%%%%%%%%%%%%%%%%%%%%%%%%%%%%%%%%%%%%%%
%%%%%%%%%%%%%% END OF INTRODUCTION %%%%%%%%%%%%%%%
%%%%%%%%%%%%%%%%%%%%%%%%%%%%%%%%%%%%%%%%%%%%%%%%%%

%-------------------------------------------------

%%%%%%%%%%%%%%%%%%%%%%%%%%%%%%%%%%%%%%%%%%%%%%%%%%
%%%%%%%%%%%%%%%%% OBSERVATIONS %%%%%%%%%%%%%%%%%%%
%%%%%%%%%%%%%%%%%%%%%%%%%%%%%%%%%%%%%%%%%%%%%%%%%%
\section{Observations}

\subsection{GNIRS/Gemini spectroscopy}
NIR spectra of \phl\ were obtained during the night of August 17, 2014, with the 8.1~m Gemini North telescope in Mauna Kea Observatory (Program ID GN-2014B-Q-29). 
We employed the Gemini Near-IR spectrograph \citep[GNIRS,][]{elias06} in the cross-dispersed mode, which allows simultaneous z+J, H and K band observations, covering the spectral range 0.8\,$-$\,2.5$\mu$m in a single exposure.
The average seeing of the night was 0.7 arcsecs. 
The instrument setup includes a 32 l/mm grating and a 0.8$\times15$ arcsec slit, giving a spectral resolution of R$\sim$1300. Individual exposures of 180~s each were taken, nodding the source in a pattern	ABBA along the slit, with a total integration time of 36 minutes. 
Right after the observation of the science frames, an A0V star was observed at a similar airmass, with the purpose of flux calibration and telluric correction.
 
\begin{figure*}
    \includegraphics[width=0.9\textwidth]{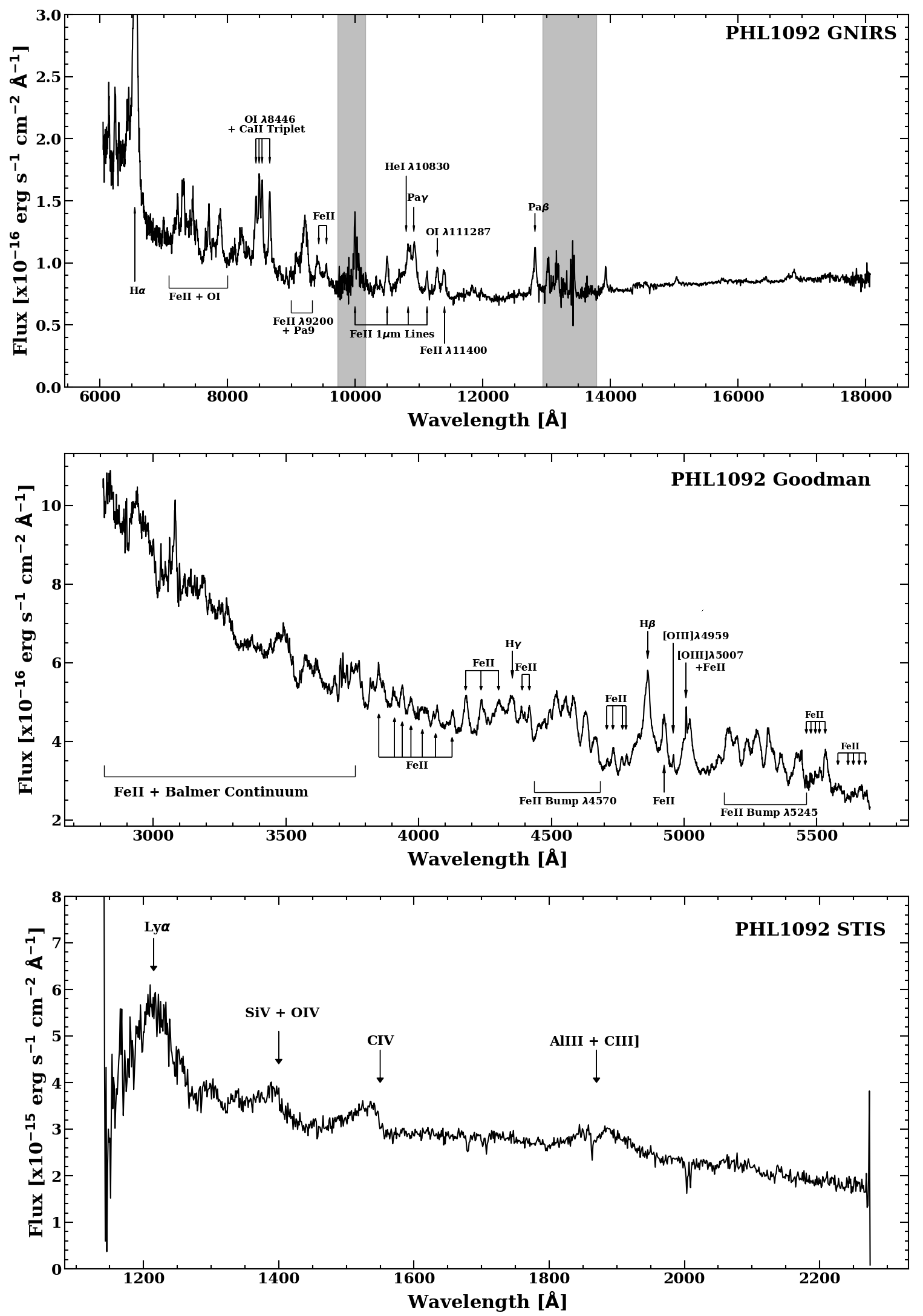}%[width=1.0\textwidth]
    \caption{Observed \phl\ spectrum with GNIRS, Goodman, and STIS. The spectra were corrected by redshift, $z=0.396$. 
    Top panel: Full reduced and flux calibrated NIR spectrum of \phl\ observed with GNIRS/Gemini North. 
    The gray shade areas mark the regions affected by telluric abortion.
    Middle panel: Fully reduced and flux calibrated optical spectrum of \phl\ observed with Goodman/SOAR. 
    Bottom panel: Ultraviolet spectrum of \phl\ as observed by STIS/HST.
    In all three panels, prominent emission lines and \feii\ multiplets are identified by the arrows.}
    \label{fig:phl_spec}
\end{figure*}

The NIR data were reduced using the XDGNIRS pipeline (v2.0)	\footnote[1]{Based on the Gemini IRAF packages}, which delivers a full reduced, wavelength and flux calibrated, 1D spectrum with all orders combined \citep[see][for a more detailed description of the software]{mason15}. 
Briefly, the pipeline cleans the 2D images from radiative events and prepares a master flat constructed from quartz IR lamps to remove pixel to pixel variation. 
Thereafter, The s-distortion solution is obtained from daytime pinholes flats and applied to the science and telluric images to rectify them. 
Arc images are used to find the wavelength dispersion solution. 
The 1D spectrum is then extracted from the combined individual exposures. 
The telluric features from the science spectrum are removed using the spectrum of a A0V standard star. 
Finally, the flux calibration was achieved assuming a black body shape for the  standard star \citep{pecaut13} scaled to its K-band magnitude \citep{skrutskie06}. 
The orders are finally combined in to a single spectrum as shown in the top panel of Figure~\ref{fig:phl_spec}.

\subsection{Goodman/SOAR spectroscopy}
Optical spectroscopy of PHL1092 was obtained on the night of December 12, 2014, with the 4.1 m Southern Observatory for Astrophysical Research (SOAR) Telescope at Cerro Pachon, Chile. 
The observations were carried out using the Goodman Spectrograph \citep{clemens04}, equipped with a 400~l/mm grating and a 0.8 arcsec slit width, giving a resolution R$\sim$1500. 
The target was observed for a total of 45 minutes in three individual exposures of 15\,minutes each. 
The standard star LTT\,1020 was observed for flux calibration. HgAr arc lamps were observed after the science frames for wavelength calibration. 
Daytime calibrations include bias and flat field images.
 
The optical data was reduced using standard IRAF tasks. 
First, the bias frames were combined and subtracted from the remaining images. 
The images were then divided by a single averaged and normalized master flatfield image. 
The wavelength calibration of the science and star frames were achieved by applying 	the dispersion solution obtained from the arc lamp frames. 
The 1D spectra of LTT~1020 were then extracted and combined to derive the sensitivity function, later applied to the \phl\ 1D spectrum. The two atmospheric bands, at 6870 and 7600\AA{}, were modeled and removed using the following procedure. First, we interpolate the continuum between the two ends of each atmospheric bands. Second, we divided the original standard star spectrum to that without the atmospheric bands. The ratio between them produce a template were the continuum is equal to one except in the regions containing the atmospheric bands. Third, We divided the \phl\ spectrum to that of the template.
The final flux calibrated optical spectrum of \phl\ is shown in the middle panel of the Figure~\ref{fig:phl_spec}.

\subsection{STIS/HST spectroscopy}
Ultraviolet spectroscopy for \phl\ was available at the Hubble Space Telescope science archive. The spectrum was taken on  August 20, 2003 using the Space Telescope imaging Spectrograph (STIS) in combination with the filter G230L and a total integration time of 5746\,s. It covers the rest frame wavelength interval 1120\,\AA{} $-$ 2240\,\AA{} with a resolution of R$\sim$600\,km\,s$^{-1}$. The bottom panel of the Figure~\ref{fig:phl_spec} shows the UV HST spectrum with the emission lines relevant to this work identified by black arrows.

Note that neither UV, optical nor NIR spectra of \phl\ were taken simultaneously. As no overlap region exists between the different data sets, no effort was made to put them into a common flux level mostly because of potential variability effects, which may produce relative shifts between the continuum levels and emission lines along the UV, optical and infrared regions. However, the spectral gap between the optical and NIR observations is only 300\,$\AA{}$ and the continuum flux at the red end of the Goodman spectrum and blue edge of the GNIRS spectrum is consistent with an underlying power-law continuum (see figure~\ref{fig:phl_spec}).
This consistency suggest no (or very small) variability between the two observations.
Assuming a conservative scenario, we consider the mean value of AGN fractional variability from \citet{kollatschny06} as representative of the variability effect in \phl. 
They estimated an uncertainty of $\sim$11$\%$ on the optical fluxes due to variability. \citet{duetal18} found a fractional variability $\lesssim$ 10 \%\ for a sample of high accretion rate AGN.
This value is also in good agreement with \citet{hu15}, who found an average fractional variability of 10$\%$ for \feii\ when compared with \hb. 
Therefore, we use this values as a variability bias on all our measurements in this work.

%%%%%%%%%%%%%%%%%%%%%%%%%%%%%%%%%%%%%%%%%%%%%%%%%%
%%%%%%%%%%%%% END OF OBSERVATIONS %%%%%%%%%%%%%%%%
%%%%%%%%%%%%%%%%%%%%%%%%%%%%%%%%%%%%%%%%%%%%%%%%%%

%-------------------------------------------------

%%%%%%%%%%%%%%%%%%%%%%%%%%%%%%%%%%%%%%%%%%%%%%%%%%
%%%%%%%%%%%%%%%%% METHODOLOGY %%%%%%%%%%%%%%%%%%%%
%%%%%%%%%%%%%%%%%%%%%%%%%%%%%%%%%%%%%%%%%%%%%%%%%%
\section{Spectral Features}

\subsection{Continuum and \feii\ Emission}

In order to measure the \feii\ content in \phl\, we first carry out a proper continuum emission subtraction. 
To this purpose, we assume that the rest-frame UV to optical continuum is represented by a power-law function.
Besides, hundreds of thousands of blended \feii\ multiplets	form a pseudo-continuum from UV to NIR \citep{sigut03}. 
Because both components are spatially unresolved, it is not possible to measure them independently. 
Moreover, in moderate to strong \feii\ emitters, it is difficult to find out spectral windows free of both continua. 
Therefore, in order to disentangle them, the best approach is to model these two components simultaneously.
    
The \feii\ emission is best represented by templates, usually derived from \izwi, the prototype  NLS1.
In the optical, \citet{boroson92} constructed an empirical template of the \feii\ from the \izwi\ spectrum by removing all other lines different from \feii\ in that AGN.  \citet{Vestergaard01} employed a similar approach to obtain an empirical UV template of this ion using a high S/N spectrum of \izwi\ observed with HST/STIS.

Other authors have constructed templates using theoretical models based on assumptions about the physical conditions in the \feii\ emission region. \citet{kovacevic10}, for instance, used the relative intensities of groups of multiplets and physical constrains to construct their template. \citet{tsuzuki06} combined the empirical approach of \citet{boroson92} and \citet{Vestergaard01} with theoretical models of the \feii\ lines using CLOUDY \citep{ferland99} to developed their templates. In the NIR, the only template available in the literature is from \citet{garcia12}. It covers the wavelength interval 8000-11600\AA\ and was constructed using a semi-empirical approach, that is, by combining \citet{sigut03} models with modifications to match the observed spectrum of \izwi. \citet{marinello16} and \citet{aldama15} already demonstrated that this template successfully reproduces the NIR \feii\ emission in several AGN.

Thus, in order to fit the optical continuum of \phl\ we will use a power-law plus the \feii\ template from \citet{boroson92}. 
Because the lines of our interest are concentrated in the region between 4000-5700\,\AA, we restrict the fit to this region.

The continuum plus \feii\ emission were fit using Equation~\ref{eq:opt_cont}:

\begin{equation}
	F(\lambda) = F^{PL}_{\lambda} + (F^{\rm FeII}_{\rm \lambda}*G_{\lambda})
	\label{eq:opt_cont}
\end{equation}

\begin{figure}
     \includegraphics[width=0.99\columnwidth]{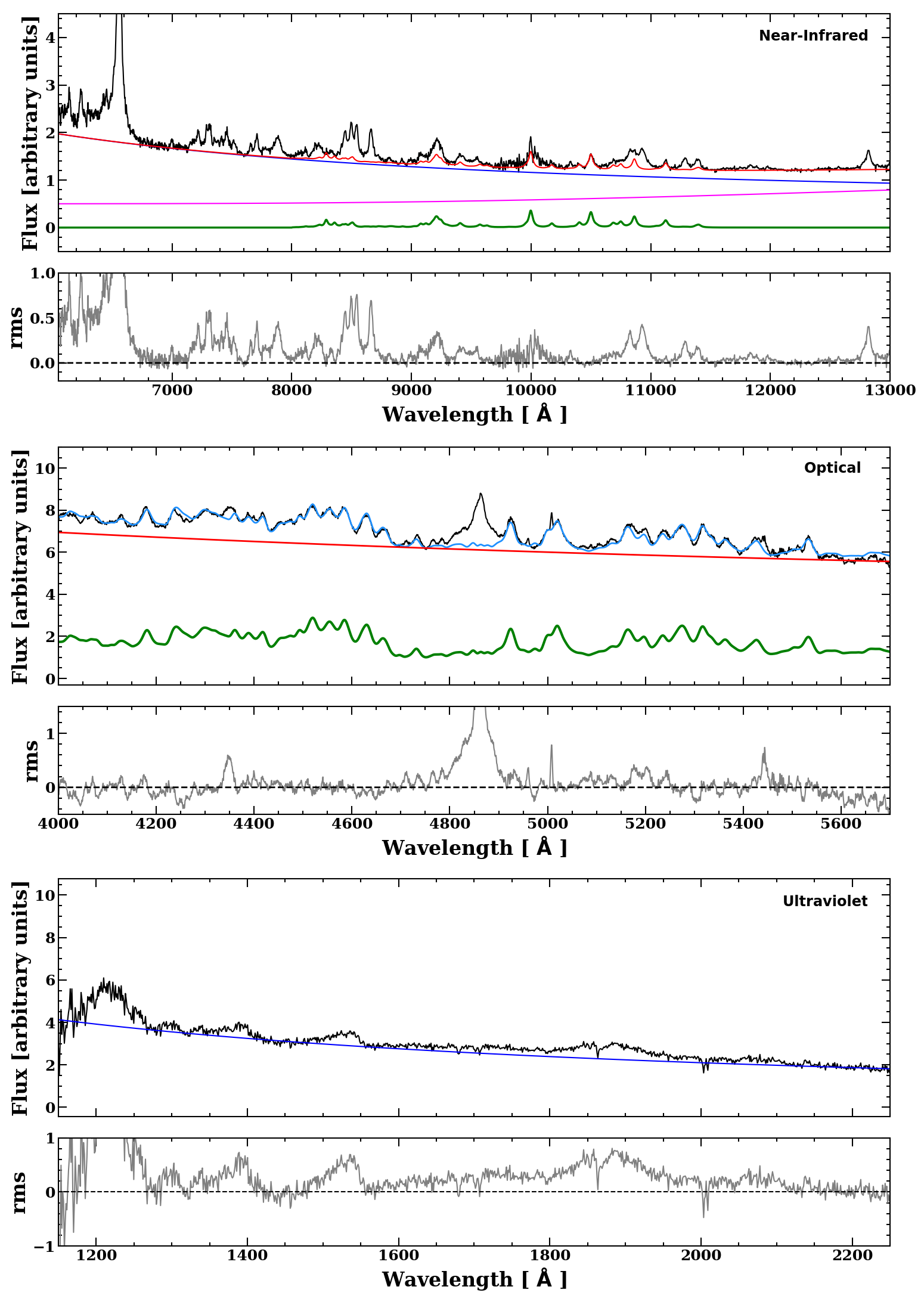}
     \caption{Continuum components in the optical and NIR spectra of PHL\,1092 (shown in black). 
     Top panel: NIR region of the spectrum. The power law, black body, and \feii\ template composing the continuum of the source are shown in blue, magenta and green lines, respectively. The sum of these components is show in red. 
     Middle panel: optical region of the spectrum. The power law and \feii\ emission are shown in red and green, respectively. The best model is show in cian.
     Bottom panel: UV region of the spectrum. Blue line shows the power law fitted to the spectrum.
     In all panels, the bottom part of the panel shows the 'pure emission line' spectrum (in grey), with the \feii\ and continuum subtracted.}
     \label{fig:contfit}
\end{figure}

Where $F^{PL}_{\lambda}\varpropto\lambda^\alpha$ describes the power law, and $F^{\rm FeII}_{\rm \lambda}$ represents the \citet{boroson92} \feii\ template convolved with a Gaussian kernel, $G_{\lambda}$, in velocity space, $V_{\rm FeII} = \sqrt{V_{\rm temp}^2+V_{\rm conv}^2}$.
To fit the continuum and the emission lines (see next section) we used our own python code, which makes use of the SCIPY library and the curve$\_$fit function. All fitting procedures in this work use this python function to estimate the best parameters in each case, obtained by minimization of the $\chi^2$ along the spectral region of the fit, masking the emission lines \hb, \oiii$\lambda\lambda$4959,5007 and H$\gamma$.
The final model, its components and the pure emission line spectrum resulting from its subtraction can be seen in middle panel of Figure~\ref{fig:contfit}. 

To model the continuum in the UV we used a single power law function, F$_{\lambda}^{PL}$. The \feii\ emission, if present, is at the continuum level and blended to the noise of the spectrum. We found a power law index of $-1.2$, consistent with the optical counterpart of the \phl\ spectrum. The bottom panel of Figure~\ref{fig:contfit} shows the modeled continuum and the pure emission lines spectrum. Prominent lines present in the spectrum are Ly$\alpha$, \civ$\lambda$1549, C\,{\sc iii}]$\lambda$1909, Al\,{\sc iii}$\lambda$1860, and Si\,{\sc iii}$\lambda$1892.

The NIR continuum were modeled using similar approach to that of the optical region. However, it was necessary to include a third component to account for the excess of continuum emission over the underlying power-law component redwards of $\sim1\mu$m, attributed to dust heated by the AGN \citep{landt08}. Thus, a three component model were used for this purpose:
(i) The same power-law observed in the optical, extended to the NIR; 
(ii) the \citet{garcia12} template for the NIR \feii\ pseudo-continuum; and
(iii) a black body function, to account for the dust emission.
This model is represented by equation~\ref{eq:nir_cont}:

\begin{equation}
	F(\lambda) = F^{PL}_{\lambda} + F^{\rm BB}_{\rm \lambda} + (F^{\rm FeII}_{\rm \lambda}*G_{\lambda})
	\label{eq:nir_cont}
\end{equation}  

where $F^{PL}_{\lambda}\varpropto\lambda^\alpha$ is the power law, $\rm F^{\rm BB}_{\rm \lambda}=B_{\lambda}(T)$ is a Planck function for the warm dust,  $F^{\rm FeII}_{\rm \lambda}$, is the \cite{garcia12} template broadened by convolving it with a kernel, $G_{\lambda}$, using as reference the width (in velocity space) of the line \feii$\lambda$10502, the strongest isolated \feii\ line in the NIR. The best fitted model, the individual components and the residual emission line spectrum are plotted in the top panel of Figure~\ref{fig:contfit}. 

The FWHM of the kernel employed to convolve the NIR template was 1150\,km\,s$^{-1}$, consistent with the value of 1200\,km\,s$^{-1}$ obtained for the optical template. Moreover, the power law index obtained for the optical and NIR are 1.26 and 1.39, respectively. Within uncertainties, they are consistent, implying that we are indeed observing in the NIR the optical extension of the continuum associated to the central source. The small difference between these two values is probably due the missing part of the spectrum, which introduces a small uncertainty in the individual slopes. Finally, the black body temperature obtained from the fit was 1290\,K, which is consistent with the temperature of warm dust found in AGN \citep{granato94,ardila06}, close to the sublimation temperature of the dust grains, $\sim$ 1600\,K.

From figure~\ref{fig:contfit}, we see that the \feii\ template suitably reproduces the optical \feii. There have been claims in the literature that a narrow component of \feii\ could also contribute to the emission in this spectral region \citep{veroncetty04,bruhweiler08,dong10}.  The models from \citet{bruhweiler08} imply that a significant part of \feii\ bump at 4570\AA\ could be produced by a narrow system of \feii\ lines. \citet{dong10}, using template modeling, found residual narrow \feii\ emission in a SDSS sample of NLS1, although much weaker than the broad component. They also noticed that narrow permitted \feii\ lines are completely absent in Seyfert 2 galaxies. We did not identify residual narrow \feii\ lines neither in the optical nor in the NIR spectrum of \phl. If this narrow component is present, it should be visible, at the very least,  in the isolated NIR \feii\ lines such as \feii$\lambda$10502 and \feii$\lambda$11127 after subtraction of the broad component. We therefore conclude that the strength of a putative contribution of a narrow component to the \feii\ spectrum should be negligible. 

Three features, though, call the attention in the NIR pure emission line spectrum: the broad emission line around $\lambda$9200, the small bump redwards of He\,{\sc i}$\lambda$10830, and a set of unidentified lines at $\lambda$11400. 

In order to search for their origin, we should recall first that the \feii\ template of \citet{garcia12} was constructed based on the models of \citet{sigut03}	with modifications in some multiplet strengths to match the spectrum of \izwi. The three features mentioned above were, in fact, modified during the construction of the template.  Since \phl\ is a super-strong \feii\ emitter, it is very likely that these features had their intensities underestimated from the template derived from \izwi. In this context, the observed excess of emission at 9200\,\AA, 10800\,\AA\ and 11400\,\AA\ is genuine and not properly modeled by the \citet{garcia12} template (see next section).  Note, however, that the residual bump redward of He\,{\sc i} could also be due to an additional broad component of that line.  If this hypothesis is correct, we should also detect a similar broad component (in FWHM and relative position from the rest-wavelength of the line) in \pab, which is not the case. We conclude that the \feii\ transitions leading to the emission lines at 9200\,\AA, 10800\,\AA\ and 11400\,\AA\ are enhanced in \phl. In order to confirm whether that excess of \feii\ emission is common in extreme \feii\ emitters, a larger sample of such objects would be necessary.

\subsection{The BLR spectrum of \phl } \label{BLRspec}

The fit described in the previous section allowed us to remove from the observed optical to NIR spectrum the continuum emission due to the central source and the \feii\ pseudo-continuum.
The pure nebular spectrum is clearly dominated by low ionization lines such as \oi, \caii, and H\,{\sc i} emitted in the BLR. 
Narrow forbidden emission lines such [Fe\,{\sc ii}]$\lambda$12570, \oiii$\lambda$4959 and \siii$\lambda$9531 are rather faint.
In the UV spectra, emission lines of  He\,{\sc ii},  \civ, Al\,{\sc iii}, S\,{\sc iii}], C\,{\sc iii}], Si\,{\sc iv} and O\,{\sc iv}] were identified.
Parameters of the broad lines (flux and width)  were derived assuming that the  line profiles can be represented by a single or a combination of Lorentzian or Gaussian functions. The best solution was obtained when the reduced $\chi^2$ of the fit reaches the minimum value. All line widths presented here were corrected for instrumental broadening using $FWHM(real)^2 = FWHM(observed)^2 + FWHM(instrumental)^2$, where FWHM(instrumental) is 360\,km\,s$^{-1}$, 180\,km\,s$^{-1}$ and 80\,km\,s$^{-1}$ for GNIRS, Goodman and STIS, respectively.

\begin{figure*}
    \includegraphics[width=1.0\textwidth]{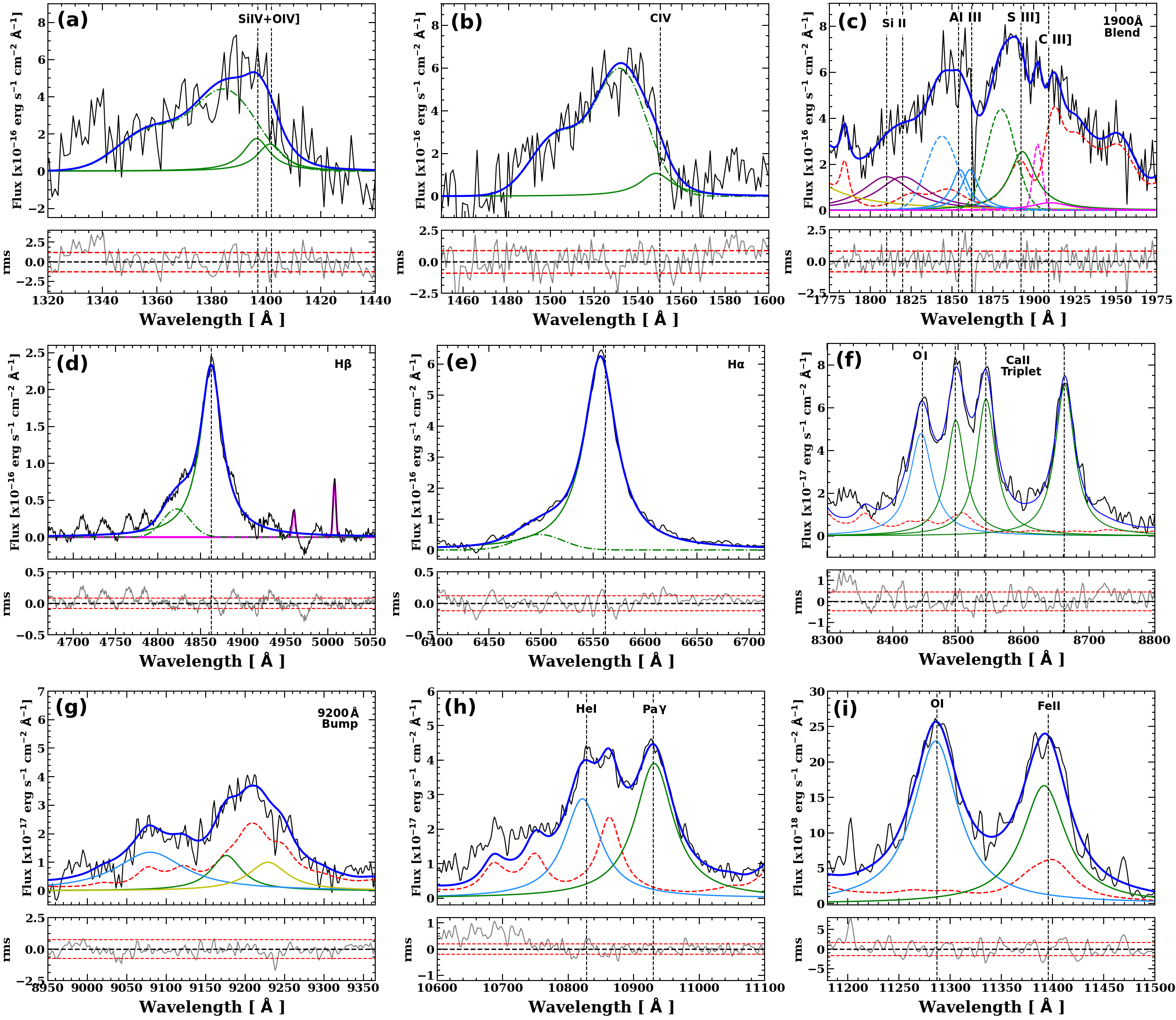}%[width=1.0\textwidth]
    \caption{Line fit of all lines of interest for this work. 
    Panel (a) shows the 1400\AA{} complex, which consist of the Si\,{\sc iv}$\lambda1397$+O\,{\sc iv}]$\lambda1402$ broad lines (solid green) and a blushifted component (dashed green).
    Panel (b) shows the \civ\ line. The broad line was fitted using a Lorentzian (solid green) added with two Gaussians to model its blueshifted component (dashed green).
    Panel (c) shows the 1900\AA{} blend. The broad components of C\,{\sc iii}]$\lambda1909$, Si\,{\sc iii}]$\lambda1982$, Al\,{\sc iii}$\lambda$1860, and S\,{\sc ii}$\lambda$1812 are denoted by the magenta, green, pale blue, and purple solid lines, respectively. The blueshifted components of each line are shown in dashed lines of same color. The yellow solid line shows part of the red wing of Ni\,{\sc III}]$\lambda1750$.
    Panel (d) shows the \hb+\oiii\ region. The \hb\ line and the blueshifted component are shown in green solid and dot-dashed lines, respectively, and the \oiii\ lines are shown in magenta.
    Panel (e) shows the \ha\ line fit. The solid and dot-dashed green lines show the broad component and the blueshifted component, respectively.
    Panel (f) plots the \oi+\caii\ triplet region. The \oi\ is shown in pale blue and the \caii\ triplet is shown in green. 
    Panel (g) shows the 9200\AA{} \feii\ bump. The solid yellow line shows the Pa9 line, and the solid green and pale blue lines shows the complementary \feii\ lines introduced to obtain the missing \feii\ flux in the bump.
    Panel (h) shows the \hei$\lambda$10830 (pale blue), the \pag\ lines (green). 
    Panel (i) shows the \oi$\lambda$11287 (solid pale blue linee) and the \feii\ feature at 11400\AA{} (solid green line.
    In all panels the bold blue line shows the complete fit of all components, observed spectrum in black and the \feii+Fe\,{\sc iii} template in dashed red.}
    \label{fig:linefit}
\end{figure*}

The UV spectrum of \phl\ has three main features that carry important information about the BLR: Si\,{\sc iv}, \civ, and the blend around 1900\,\AA, formed by the contribution of different species (see below). Since \phl\ has a strong \feii\ emission and narrow broad lines, it is classified as a population ``A" AGN. In such type of sources, the broad lines profiles are better represented by a Lorentzian function. 
The emission line at 1400\AA{} is actually a blend of Si\,{\sc iv}$\lambda1397+$O\,{\sc iv}]$\lambda1402$ \citep{aldama18}.
We fit that feature using two Lorentzians to represent the rest-frame component of the blend, and a combination of two Gaussians profiles to represent the blue asymmetry observed in the line. The \civ\ line were fit using a Lorentzian with the addition of two Gaussians to account for the blue asymmetry, usually associated to outflows \citep{coatman16}. This approach in both fits warrants a proper modeling of the asymmetry without $ad-hoc$ assumptions about its origin. The flux and the FWHM of these lines are listed in Columns 2 and 3, respectively, of Table~\ref{tab:linefit} and the best fit with the individual components can be observed in Figure~\ref{fig:linefit}\,(a) and (b). Note that the broad components of \civ, Si\,{\sc iv}, and O\,{\sc iv}] are especially uncertain due to the much stronger emission of the Blue component. For that reason, the fluxes associated to the broad components of these lines are marked with a ``:". We also measured the specific flux in a small range at $\approx$ 1400 \AA\ and at 1549 \AA\ where we expect the rest frame component emission of Si\,{\sc iv}, and C\,{\sc iv}), respectively. We obtain $5 \cdot 10^{-16}$ erg s$^{-1}$ cm$^{-2}$ \AA$^{-1}$ for Si\,{\sc iv} and $3 \cdot 10^{-16}$ erg s$^{-1}$ cm$^{-2}$ \AA$^{-1}$ for C\,{\sc iv}. The ratio  Si\,{\sc iv} / C\,{\sc iv} is therefore $\approx 1.7$, consistent with the ratio reported in Table \ref{tab:lineratio}.

The blend at 1900\,\AA\ is composed of \feii, Fe\,{\sc iii}, C\,{\sc iii}]\,$\lambda$1909, Si\,{\sc iii}]\,$\lambda$1892, and Al\,{\sc iii\,}$\lambda$1860. We follow the procedure outlined in \citet{aldama18} to fit this bump. First, we used the \citet{Vestergaard01} UV \feii+Fe\,{\sc iii} template with  Lorentzians profiles to model the last three lines. Note that Al\,{\sc iii}$\lambda$1860 is actually a doublet at $\lambda\lambda$1854,1862, with equal intensity (1:1). We found that Si\,{\sc iii}] and C\,{\sc iii}] has a prominent blueshifted component, which according to \citet{aldama18} is rare although observed before in even more extreme regimes than that observed in \phl\ \citep{aldama18b}. We modeled the blueshifted components with a Gaussian profile. The corresponding FWHM, fluxes and line shifts are listed in Table~\ref{tab:linefit}. The best fit can be observed in Figure~\ref{fig:linefit}\,(c). Table \ref{tab:lineratio} provides UV line ratios relevant to the classification (PHL 1092 meets the ``extreme Population A'' criteria of \citealt{marziani14}), and to the tentative definition of the BLR physical conditions (Sect. \ref{physblr}).

We fit \ha\ and \hb\ with  Lorentzian profiles, representing the BLR contribution. In addition, we noticed a blue asymmetry in these lines, which was fit with a Gaussian component, following the approach of \citet{negrete18}. Moreover, we employed one Gaussian to model each of the \oiii$\lambda\lambda$4959,5007 doublet. We did not find evidence of the \nii\ doublet around \ha. Note that the \siii$\lambda$9531, usually the strongest NIR narrow forbidden line \citep{landt08,riffel06,mason15}, was not detected in our spectrum, suggesting that the NLR contribution is likely below the detection limit of the spectrograph. 
The lack of \nii\ and [S\,{\sc ii}] detection   is consistent with the weakness of \siii\ and \oiii. Since the \oiii\ and \siii\ are usually the strongest narrow lines in the optical and NIR, respectively, and [N\,{\sc ii}] and [S\,{\sc ii}] are just a fraction of \oiii\ ($\sim0.3$), it is not expected that they show up in the spectrum.
 The panels  (d) and (e) of Figure~\ref{fig:linefit} show the best fit for \hb\ and \ha, respectively. The parameters of the fit are listed in Table~\ref{tab:linefit}. Moreover, the \feii\ emission bump centered in $\lambda$4570 was measured using the template fit in the previous section. The flux labeled \feii\ (4570 Bump) presented in Table~\ref{tab:linefit} is the integrated flux of the bump in the wavelength interval 4434--4684\AA\ \citep{boroson92}.

\begin{table*}
    \centering
    \caption{Emission line fitting results.}
    \label{tab:linefit}
    \begin{tabular}{lcccc} % four columns, alignment for each
      \hline
      \hline
      Line         					& Flux       					& FWHM          & Integrated &  Line shift$^a$ \\  
      								&(x10$^{-16}$\,erg\,s$^{-1}$)	&(km\,s$^{-1}$) & S/N 		  &  (km\,s$^{-1}$) \\
    \hline
      H$\beta\,_{\rm BC}$			&112.01$\pm$2.43 	&1850$\pm$100 	&46	& 0\\
      H$\beta\,_{\rm Blue}$	  	    &21.85$\pm$1.85 	&2300$\pm$200 	&8  & -2470\\
      \oiii\,$\lambda$4959 			&1.58$\pm$0.30  	&300$\pm$50		&3	& 0\\
      \oiii\,$\lambda$5007 			&3.13$\pm$0.35		&300$\pm$50		&5	& 0\\
      H$\alpha\,_{\rm BC}$			&363.58$\pm$4.15	&1715$\pm$100	&116& -218\\
      H$\alpha\,_{\rm Blue}$        &28.11$\pm$1.18     &2350$\pm$210   &6  & -2700\\
      \oi\,$\lambda$8446			&31.26$\pm$1.90		&1350$\pm$100	&23	& -95\\
      \caii\,$\lambda$8495			&29.47$\pm$1.85		&1250$\pm$120	&23	& 0\\
      \caii\,$\lambda$8543			&35.12$\pm$1.84		&1250$\pm$120	&29	& 0\\
      \caii\,$\lambda$8662			&29.23$\pm$1.90		&1250$\pm$120	&31	& 0\\
	  \feii\,$\lambda$9998			&24.90$\pm$2.20		&1150$\pm$75 	&--	& 0\\      
      \feii\,$\lambda$10502			&23.14$\pm$0.90		&1150$\pm$75 	&25	& 0\\
      \hei$\,_{\rm BC}\lambda$10829	&30.25$\pm$1.50  	&1860$\pm$150	&26	& -219\\
      \feii\,$\lambda$10863			&22.315$\pm$0.90  	&1150$\pm$75 	&23	& 0\\
      \pag$\,_{\rm BC}$				&43.53$\pm$2.00  	&1900$\pm$130	&35	& 0\\
      \feii\,$\lambda$11127			&13.10$\pm$1.12  	&1150$\pm$75 	&20	& 0\\
      \oi\,$\lambda$11287			&19.06$\pm$1.45  	&1350$\pm$100	&23	& -55\\
      \feii\,$\lambda$11400			&18.21$\pm$1.10 	&1150$\pm$75 	&23	& -105\\
      Si\,{\sc iv}\,$\lambda$1397	&32:		&2500$\pm$150	&3	& -101\\
      O\,{\sc iv}]\,$\lambda$1402	&27:		&2500$\pm$150	&3	& -106\\
      Si\,{\sc iv}+O\,{\sc iv}]$_{\rm Blue}$     &186.84$\pm$17.15 	&9400$\pm$1030	&17	& -4700\\
      C\,{\sc iv}\,$\lambda$1549    &30:		&3800$\pm$175	&4	& -184\\
      C\,{\sc iv}$_{\rm Blue}$      &255.11$\pm$22.93 	&5350$\pm$295	&10	& -3500\\
      C\,{\sc iii}]\,$\lambda$1909	&14.51$\pm$1.02  	&4200$\pm$315	&3	& 0\\
      C\,{\sc iii}]$_{\rm Blue}$	&29.35$\pm$1.78  	&1800$\pm$115	&7	& -1100\\
      Si\,{\sc iii}]\,$\lambda$1892	&76.64$\pm$5.29  	&3100$\pm$150	&12	& 0\\
      Si\,{\sc iii}]$_{\rm Blue}$	&141.25$\pm$12.67  	&3400$\pm$160	&23	& -1750\\
      Al\,{\sc iii}\,$\lambda$1860$^b$	&70.06$\pm$1.12  	&2150$\pm$175	&9	& 0\\
      Al\,{\sc iii}$_{\rm Blue}$	&111.27$\pm$17.12  	&3700$\pm$175	&20	& -950\\
    \hline				
    \feii\ (4570 Bump)				&288.55$\pm$10.3	&--		&--	\\
    \feii\ (9200 Bump)				&79.50$\pm$1.12		&--		&--	\\
    \feii\ (1\,$\mu$m lines)		&79.53$\pm$1.75		&--		&--	\\
    \hline
    \hline
    \multicolumn{2}{l}{$^a$ Negative values are regarded as blueshifts}\\
    \multicolumn{2}{l}{$^b$ Summed flux of the lines in the doublet}\\
    \end{tabular} 
\end{table*}

\begin{table*}
    \centering
    \caption{Broad emission line ratios.}
    \label{tab:lineratio}
    \begin{tabular}{lc} % four columns, alignment for each
      \hline
      \hline
      Line Ratio         					& Value \\  
    \hline
    R$_{\rm 4570}$					&2.58$\pm$0.13	\\
    R$_{\rm 9200}$					&1.02$\pm$0.10	\\
    R$_{\rm 1\mu m}$				&1.02$\pm$0.11	\\
    \hline
    Al\,{\sc iii}/Si\,{\sc iii}]	&0.91$\pm$0.09	\\
    C\,{\sc iii}]/Si\,{\sc iii}]	&0.19$\pm$0.07	\\
    \hline
    Si\,{\sc iv}+O\,{\sc iv}]/Si\,{\sc iii}]		&0.76:	\\
    C\,{\sc iv}/Al\,{\sc iii}		&0.42:	\\
    C\,{\sc iv}/Si\,{\sc iii}]		&0.39:	\\
%    Si\,{\sc iv}+O\,{\sc iv}]/C\,{\sc iv}		&1.94$\pm$0.18	\\    
    Si\,{\sc iv}+O\,{\sc iv}]/C\,{\sc iv}		&1.94:	\\ 
    
     \hline    
    (Si\,{\sc iv}+O\,{\sc iv}])/C\,{\sc iv}$^a$		&0.86$\pm$0.09	\\    
    \hline
    \hline
      \multicolumn{2}{l}{$^a$ Ratio between the summed flux of BC and Blue. }\\
    \end{tabular}
  \end{table*}
%\newpage
The \oi~$\lambda$8446 and the \caii\ triplet ($\lambda$8496, $\lambda$8542, and $\lambda$8663) were fitted considering two constrains.
First, \oi~$\lambda$8446 was constrained to have the same FWHM (in velocity space) as \oi~$\lambda$11297. Second, \caii\,$\lambda\lambda$8496,8542 and $\lambda$8663 were constrained to have the same width \citep{ardila02}. The values obtained are listed in Table~\ref{tab:linefit} and the fit is presented in Figure~\ref{fig:linefit}(f).

The bump at 9200\AA{} is a blend of \feii\ and Pa9. \citet{marinello16} showed that the NIR \feii\ template usually underestimates the total flux in that feature.
The residual emission left after subtraction of the template is composed of the residual contribution of \feii\ and the Pa9 line.
In order to obtain a more accurate measurement of the \feii\ emission in the bump we followed the technique presented by \citet{marinello16}, which consists of subtracting the Pa9 contribution to the bump after scaling down \pag\ by the theoretical \pag/Pa9 ratio (Case B).
After this subtraction, the residual should consist of a pure \feii\ emission, which can be fit with a proper function (Lorentzian in this case).
The total flux of the bump is the sum of the fluxes from the template added to that the residual (see  Table~\ref{tab:linefit}).  Figure~\ref{fig:linefit}(g) shows the fit applied to this region.

The only prominent Paschen line detected in the spectrum is \pag, which is moderately blended with \hei$\lambda$10830. Other Paschen lines fall in the region of poor atmospheric transmission.
We model the above two lines with a Lorentzian profile, representing the broad component.
No narrow component was detected above 3$\sigma$ level.
The FWHM of the broad components of \pag\ and \hei\ are 1900 and 1850\,km/s, respectively.
The value found for \pag\ is consistent to that measured in \hb, showing that the results found are robust.
The best fitted profiles can be seen in panel (h) of Figure~\ref{fig:linefit}.

The \oi$\lambda$11297 and \feii$\lambda$11400 were also fitted using a Lorentzian function. Both features are slightly blended.  
This function most suitably represents the shape of these lines compared to a Gaussian function and is consistent with the profile fit to \hb.
Note that \feii$\lambda$11400 was subtracted using the \feii\ template but it was significantly underestimated by it, leaving a strong residual emission feature.
We estimate the total flux in \feii$\lambda$11400 by summing the flux value obtained from the template and that of the additional component fit to the residuals.
Figure~\ref{fig:linefit}(i) shows the fit to \oi$\lambda$11287 and \feii$\lambda$11400.
The integrated flux and FWHM are also presented in Table~\ref{tab:linefit}.

Another important \feii\ features in the NIR are the so-called 1\,$\mu$m lines, e.g., \feii$\lambda$9998, 10502, 10863 and 11127~\AA. We measured the flux of these lines directly from the fit carried out using the \citet{garcia12} template. The results for each individual line is presented in Table~\ref{tab:linefit}. For all lines measured in this work we list in the same table the integrated signal-to-noise (S/N) and the line shifts with respect to the systemic velocity of the galaxy in Columns 4 and 5, respectively.

%%%%%%%%%%%%%%%%%%%%%%%%%%%%%%%%%%%%%%%%%%%%%%%%%%
%%%%%%%%%%%%%% END OF METHODOLOGY %%%%%%%%%%%%%%%%
%%%%%%%%%%%%%%%%%%%%%%%%%%%%%%%%%%%%%%%%%%%%%%%%%%

%-------------------------------------------------

%%%%%%%%%%%%%%%%%%%%%%%%%%%%%%%%%%%%%%%%%%%%%%%%%%
%%%%%%%%%%%%%%%%%% DISCUSSION %%%%%%%%%%%%%%%%%%%%
%%%%%%%%%%%%%%%%%%%%%%%%%%%%%%%%%%%%%%%%%%%%%%%%%%
\section{Discussion}

\subsection{\phl\ in the Eigenvector 1 context}

It has been 27 years since \citet{boroson92} presented their seminal work that shows that most of the spectral features of an AGN can be parameterized in terms of two variables, called Eigenvector 1 (E1).
They showed that the anti-correlation between the flux ratio \feii$_{\rm 4570}$/\hb\ and the EW of \oiii\ forms an optimal parametric space where several other properties relates with.
This concept was later extended by \citet{sulentic00} to a four-dimensional parameter space, 4DE1, where the FWHM(\hb), \rop, the shift of the peak of C\,{\sc iv} relative to the systemic velocity, and the photon spectral index of the soft-Xray power-law $\Sigma$, were added.
The main physical driver of the E1 has been attributed to variations of the accretion rate among different AGN \citep{boroson02,marziani01,shen14} while the spread of the FWHM(\hb) in the optical is probably related to orientation effects \citep{shen14}. 

In order to include \phl\ in the E1 diagram it is necessary to estimate its corresponding \rop.
Using the values measured in the previously section (see Table~\ref{tab:linefit}), we estimate \rop=2.58.
The \feii\ emission in AGN can be divided into basically three categories.
Weak \feii\ emitters, with \rop\ $<$ 1, strong \feii\ emitters, with 1$<$\rop$<$2, and super-strong \feii\ emitters, with \rop$>$2.
Typical AGN are weak \feii\ emitters, with $\sim90\%$ of them in this category, and highly concentrate around \rop$\sim~0.6-0.8$ \citep{shen14}.
Strong \feii\ emitters are less common, occurring in about 5$\%$ of the AGN population \citep{lawrence88}. Super-strong \feii\ emitters are even more rare, roughly an order of magnitude less \citep{moran96,lipari93}.
In this scheme, \phl\ is in the category of super-strong \feii\ emitters.

\citet{bergeron80} were the first to identify \phl\ as a super-strong \feii\ emitter.
They showed that the optical spectrum was dominated by intense \feii\ lines and that such emission could only arise from gas of high electronic density, n$_e\sim10^{12}$~cm$^{-3}$, and low temperature, T$\sim10000$~K.
However, the value of \rop\ they reported was surprisingly high, \rop=6.2, making \phl\ the strongest \feii\ emitter ever identified.
\citet{lawrence97} proposed that values of \rop\ in the interval 4-8, published in the literature by \citet{joly91}, could be too high. They indeed found for \phl\ a \rop\ of 1.8.

\begin{figure}
     \includegraphics[width=1\columnwidth]{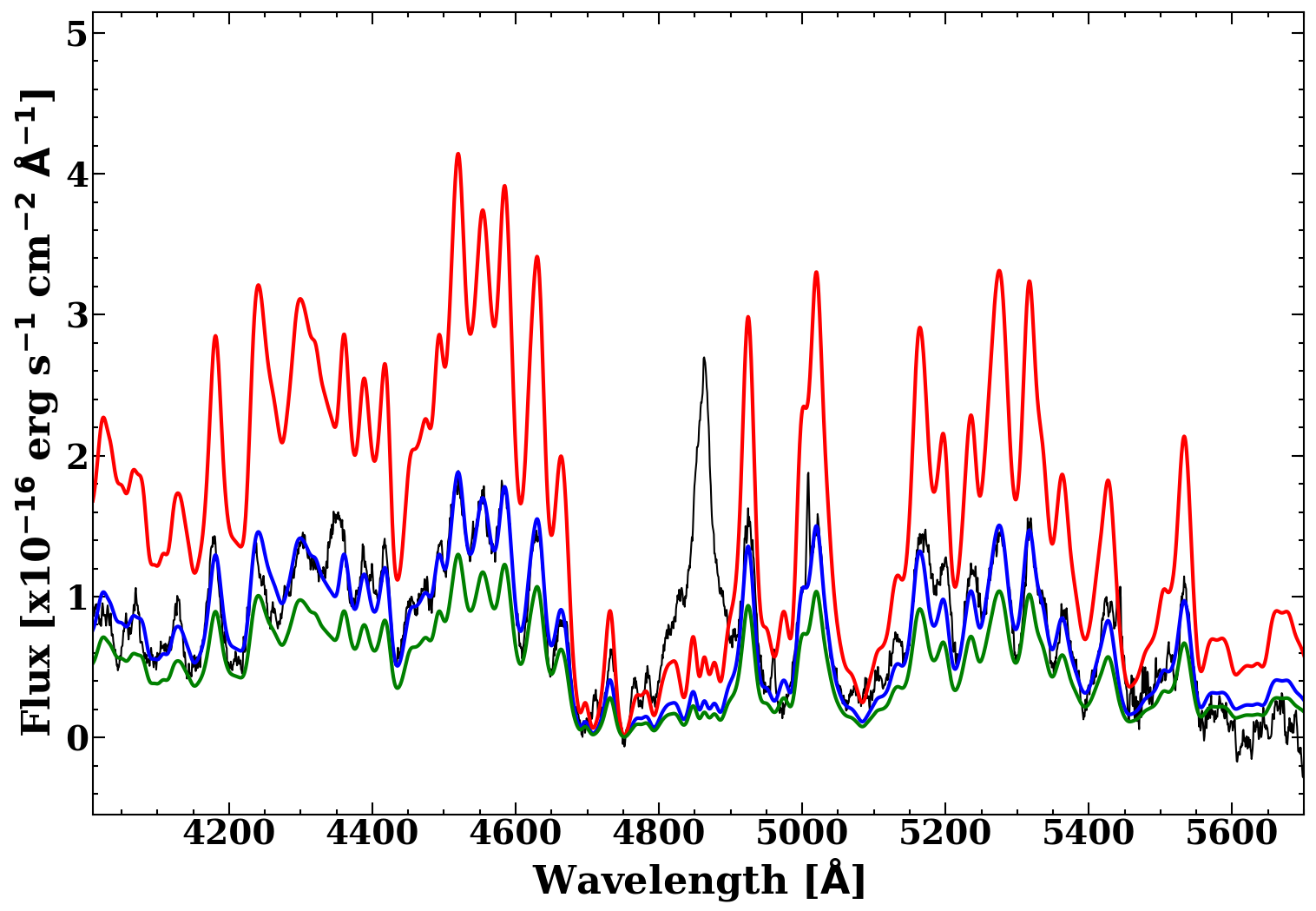}
     \caption{\phl\ with the \feii\ template scaled to the different values of \rop\  from the literature. 
     The observed spectrum with the power law subtracted is shown in black. 
     The best fitted \feii\ template in this work are shown in blue.
     The red and green lines lines show the template scaled to reproduce the \rop\ obtained by \citet{bergeron80}, \rop$=6.2$, and by \citet{miniutti12}, \rop$=1.8$, respectively.}
     \label{fig:feiicomp}
\end{figure}

The large discrepancies in the value of \rop\ found here and those in the literature  are likely due to the method employed to measure the flux contained in the \feii\ bump.
In \citet{bergeron80}, most of the flux redwards of \hb\ was associated to \feii\ multiplets, reducing the amount of \hb\ flux in the broad Lorentzian wings of the line profile and thereby increasing \rop\ to higher values.
The continuum level set in the measurements is also a factor that can lead to a smaller value of \rop.
Figure~\ref{fig:contfit} shows that a flat continuum under the \feii\ $\lambda$4570 bump would underestimate the integrated flux of that feature.
The value of \rop\ presented here is based on a simultaneous fit to the continuum and the \feii\ spectrum observed in the UV/optical region,  resulting in more robust approach.
In order to compare the different values reported in the literature with ours, we show in Figure~\ref{fig:feiicomp} our spectrum and template, and how it would look if the \rop\ were 1.8 and 6.2.
As shown in Figure~\ref{fig:feiicomp}, a \rop\ of 6.2 is unrealistic, being the value of \rop = 2.58 far more suitable to the observations.

\begin{figure}
     \includegraphics[width=1\columnwidth]{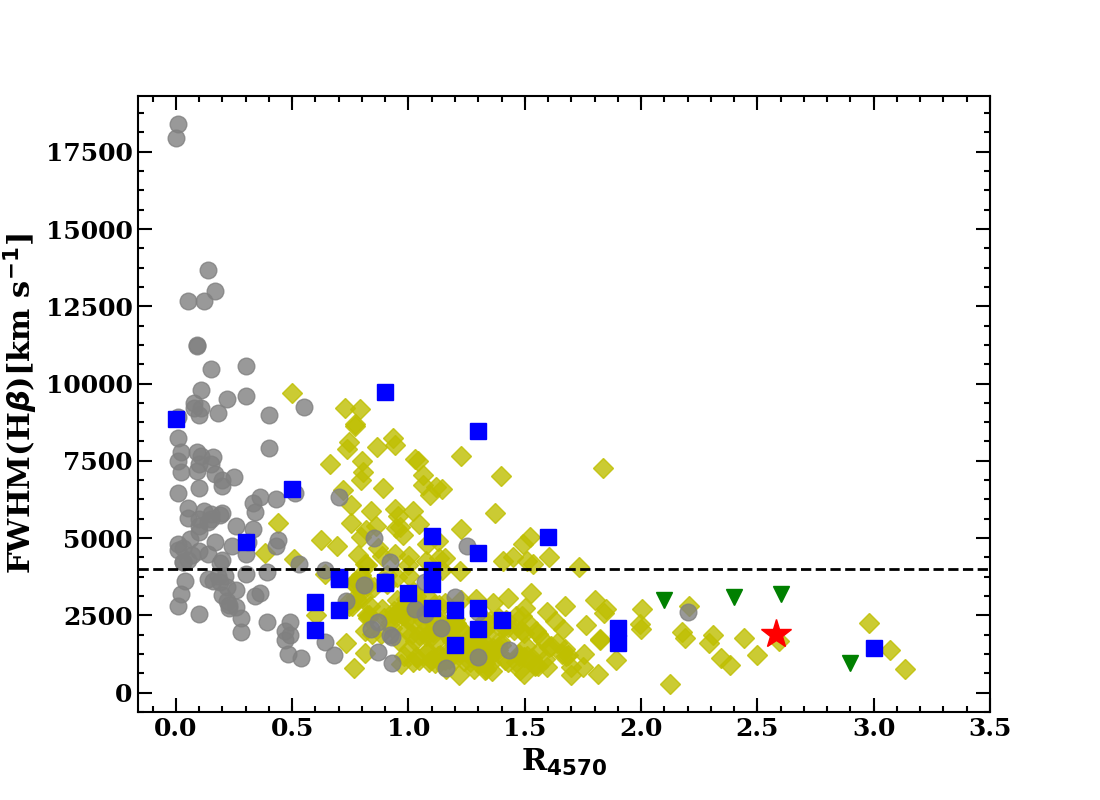}
     \caption{Eigenvector 1 optical plane. The red star shows \phl.
     The other super-strong \feii\ emitters from \citet{lipari93} are plotted in green triangles. High Eddington ratio AGN from \citet{negrete18} are denoted by yellow diamonds.
     The sources which had their \rop\ re-estimated by \citet{sniegowska18} are shown in blue squares.
     The grey circles show the sources in \citet{sulentic07}. The black dashed line show the separation
     in FWHM(\hb) of 'Population A' and 'Population B'.}
     \label{fig:ev1}
\end{figure}

The other axis of the E1 plane is the FWHM of \hb, FWHM(H$\beta$). The value obtained for the broad component is 1850\,km\,s$^{-1}$, consistent with the classification of \phl\ as a NLS1 AGN. It also  places \phl\ in the ``population A" region of the E1 optical plane \citep{marziani01}, as shown in Figure~\ref{fig:ev1}. 
In that plot, grey points represent the sources from \citet{sulentic07}, green triangles are those of \citet{lipari93} and blue squares are AGN are from \citet{sniegowska18}.
\phl\ is represented by the red star.
Even after reducing the estimate of \rop\ for \phl, it still remains as outstanding source.

Other objects with super-strong \feii\ emission were presented by \citet{lipari93}: IRA07598+6508, Mrk\,231, Mrk\,507, and IRAS18508-7815. All of then with \rop$>2$.
\citet{shen14}, analyzing the physical drivers of EV1, derived the same diagram for all SDSS (DR7) sources from \citet{shen11}. An immediate result from their work is the lack of sources with \rop$>$3.
In fact, in their E1 diagram, it is possible to see an upper \rop\ cutoff around \rop$\sim2.2$.
\citet{sniegowska18} re-analyzed 27 of the strongest \feii\ emitters (\rop$>1.3$) from the \citet{shen11} catalog. Their results show that many sources in \citet{shen11} had their \rop\ overestimated, very likely due to the automated procedure employed to derived this quantity and the low S/N of some of the spectra.
That is the case of SDSS125343.71+122721.5, which has \rop=1.75 in \citet{shen11} but is reported with negligible \feii\ emission in \citet{sniegowska18}.
They also identified another outstanding source of \feii\ emission, SDSS125343.71+122721.5.
This AGN has a \rop=2.12 in \citet{shen11} catalog and after \citet{sniegowska18} analysis, they presented two different values: 
(i) \rop=3.0 and FWHM(\hb)=1445\,km\,s$^{-1}$ for a Gaussian profile to fit \hb, and 
(ii) \rop=2.12 and FWHM(\hb)=936\,km\,s$^{-1}$ using a Lorentzian profile instead.

Similarly, \citet{negrete18} analysed a sub-sample of 302 AGN up to redshift 0.8 and Eddington ratio close to 1, extracted from the much larger sample of \citet{shen11}.
In their work they found that these sources are characterized by strong \feii\ emission. In particular, it was found that 16 sources have \rop$>2$, with 5 of them displaying \rop$>2.5$.

The higher values of \rop\ are sometimes due to the very narrow profile of \hb.
In these extreme NLS1 galaxies, the \hb\ line is dominated by a narrow component. The distinction between the contribution from the BLR and NLR are extremely subtle and, in some cases, can only be fully evaluated by means of multi-wavelength spectra (i.e. optical and NIR data).
Therefore, a careful fit to the H$\beta$ region, taking into account the power-law continuum and simultaneously the \feii\ emission, is crucial to derive consistent values of R4570. 

\citet{zamanov02} pointed out that in Population A of AGN, strong \feii\ emitters with narrow \hb\ profiles tend to have a large blueshifted component in \oiii. 
This trend, though, is not observed  in \phl.
This may be due to an intrinsic effect caused by the low strength of the \oiii\ lines.
This is in agreement with \citet{boroson92} and \citet{shen14}, who found a negative correlation between the intensity of \oiii\ and the strength of the \feii. As the blueshifted \oiii\ component is usually weaker than the main narrow component, if the former is present, it is probably below our detection limit or heavily blended with the adjacent \feii\ features. A higher spectral resolution spectrum is indeed necessary to uncover the presence of such outflow component.

Another implication of the E1 is that strong \feii\ emitters should host low black hole masses and high Eddington ratio \citep{negrete18}. 
In order to estimate the black hole mass in \phl, $M_{BH}$, we used Vestergaard \& Peterson's (2006) single epoch black hole mass equation:

\begin{equation} 
	log(M_{BH}) = log\bigg\{ \bigg[ \frac{FWHM(H\beta)}{10^3\,km\,s^{-1}}\bigg]^2 \bigg[ \frac{\lambda L_{\lambda}(5100\AA)}{10^{44}\,erg\,s^{-1}} \bigg]^{0.5} \bigg\} + (6.91\pm0.02) 
	\label{eq:bh} \end{equation}   
	
%Flux @5100
%5.099189777702589708e+03 2.920027949854417214e-16 %Dl = 2070 Mpc %Mpc2cm =
%3.08568025 * pow(10,24) %L = 4*pi*(Dl^2)*F/(1+z) 

For \phl\ we measured a FWHM(\hb)=1850\,km\,s$^{-1}$ and $L_\lambda(5100\AA{})=1.497\times10^{41}L_{\odot}$. 
It translates using Eq.\ref{eq:bh} in a black hole mass of log($M_{BH}$)=7.89$M_{\odot}$.  
Other works in the literature report different values of $M_{BH}$  for PHL1092. 
For instance, \citet{dasgupta04} using optical scaling relation for the BLR radius and single epoch method found $log(M_{BH})=8.20M_{\odot}$. 
\citet{czerny01} found $log(M_{BH})=6.09$\msol\ using X-ray variability and $log(M_{BH})=8.26$\msol\ from accretion disk fitting method. 
\citet{nikolajuk09} found $log(M_{BH})=8.46M_{\odot}$ using more recent scaling relations for the optical single epoch recipe. 
By the same method, \citet{miniutti12} found $log(M_{BH})\sim8.48M_{\odot}$. From x-ray observations, and considering a non-spinning BH \citet{miniutti12} found a $M_{BH}=8.38M_{\odot}$, consistent with their optical measurement.
Our result is only factor 1.2 lower than the average of previous values  presented in literature. 
However, note that they estimate the $M_{BH}$ using different methods and scaling relations. 
Moreover, the black hole mass obtained here for \phl\ is consistent with the average of masses obtained by \citet{rakshit17} and \citet{negrete18}. 
The Eddington ratio is defined by the ratio between the bolometric luminosity and the Eddington limit, e.g., $L/L_{Edd}=L_{bol}/L_{Edd}$, where $L_{Edd}=1.5\times10^{38}(M_{BH}/M_{\odot})$ \citep{netzer10}. 
The bolometric luminosity can be obtained applying a correction factor to the optical luminosity measured from the continuum at 5100\AA{}, $L_{bol}=c\times \lambda F_{\lambda}(5100$\AA{}), where c=7 \citep{netzer07}.
We estimate a $log(L_{bol})$=45.72 erg\,s$^{-1}$ and $L/L_{Edd}$=1.24. 
\citet{negrete18} using a sample of 334 high Eddington ratio AGN found that strong \feii\ emitters (\rop$>1$) are associated with high $L/L_{Edd}$.  Our results presented above show that \phl\ follows this same trend, having a high accretion rate and $M_{BH}$ typical of Population A AGN. 
Despite the number of sources with \rop$>2$ is much smaller than weaker \feii\ emitters, those analyzed by \citet{negrete18} and here suggest that indeed these characteristic are common to all extreme Population A AGN.

\subsection{The Fe\,{\sc ii} Emission Region}

The \feii\ emission arises from clouds located in the broad line region, which is unresolved even for closest AGN at sub-arsec resolution observations.
For this reason, the structure and location of the gas where these lines are produced must be derived using integrated spectra. Better estimates of the BLR structure can be achieved using reverberation mapping. Indeed, studies focused on variability have detected \feii\ time lags for a few AGN \citep{rafter13,barth13,du16}.
The main result gathered from these works is that the clouds emitting \feii\ are located at distances that coincide with that of the \hb\ emission region and up to twice that from the central source. However, no consensus regarding the precise location of the \feii\ emission region has been achieved.
%A consistent picture about the \feii\ emission region is subjected to large uncertainties. 
%A comparative analysis of the broad lines from different ions may help to put additional constrains to that discussion.

\begin{figure*}
    \includegraphics[width=1.0\textwidth]{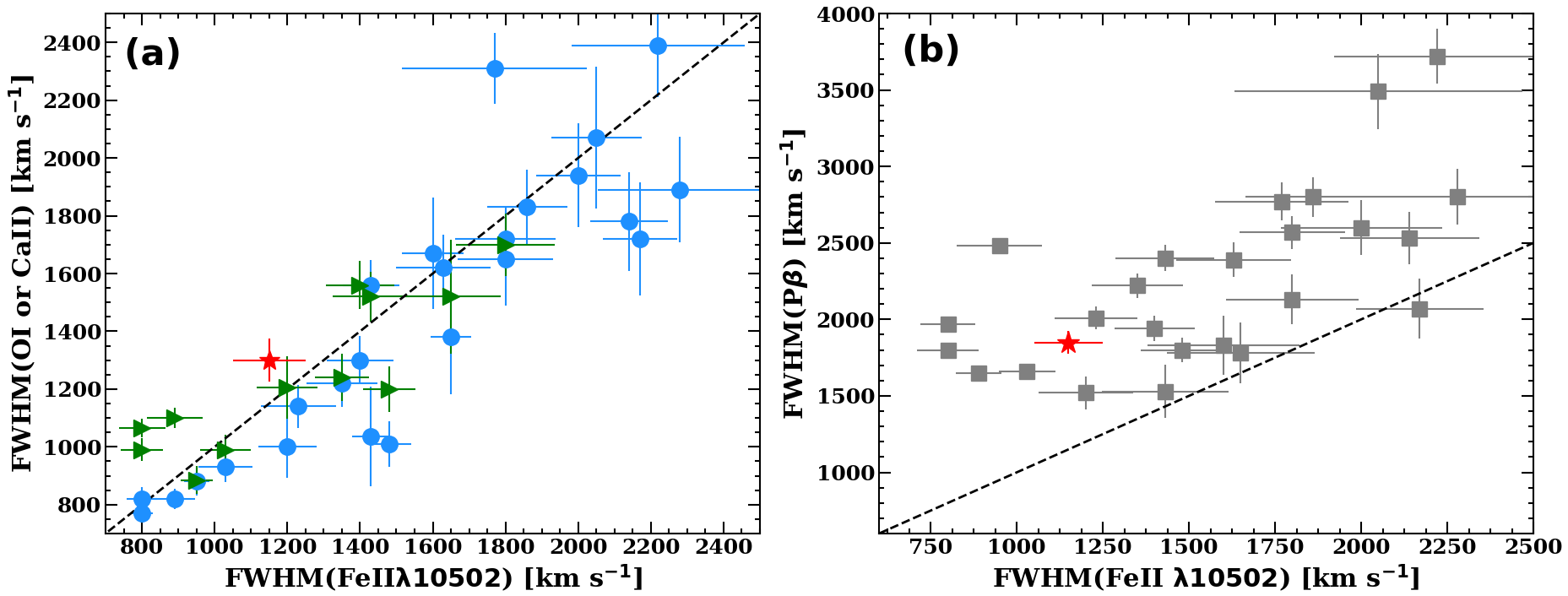}
    \caption{Correlation between the FWHM of the low ionization lines. 
        Panel (a): FWHM(\oi\,$\lambda11127$) (blue squares) and FWHM(\caii\,$\lambda8662$) (green triangles) plotted in 
    against of the FWHM(\feii\,$\lambda10502$), the most prominent isolated \feii\ line, using the sample of \citet{marinello16}. 
    Panel (b): FWHM(\pab) against of FWHM(\feii\,$\lambda10502$) is plotted in grey squares for the \citet{marinello16} sample. 
    In both panels the black dashed line is the identity function and the \phl\ is denoted by the red star.}
    \label{fig:fwhm}
\end{figure*}

In order to get clues about the possible location of the \feii\ emitting region in \phl\ we plot in Figure~\ref{fig:fwhm}(a) the FWHM of \feii\ versus the FWHM of \oi\ and \caii\ from the sample of \citet{marinello16}. We add the value of \phl\ as a red star for comparison.
The results show that the FWHM of these lines follow a similar distribution, being slightly scattered around the identity line.
Under the virial assumption, the relative values of the line widths can be used as a proxy of the distance ratios from the central source, providing a relative location of the emitting region if variations in FWHM for different ions within the same object are detected.
In this context, Figure~\ref{fig:fwhm}(a) shows that \feii, \oi, and \caii\ in \phl\ are likely formed in gas that is co-spatial, at the same distance from the AGN.

Similar results were already reported in the literature.
For instance, \citet{ardila02} studying the physical process behind the \oi\ emission lines found that \feii\ and \oi\ lines share the same profile shape, and very similar FWHM.
Using a similar argument, \citet{matsuoka07} and  \citet{aldama15} employed the \oi\ emission line to probe the physical conditions of the emitting region of \oi\ and \feii.
They all agree that these lines are formed in a same portion of the BLR, in dense clouds ($\rm n_H>10^{11.5}$~cm$^{-3}$) illuminated by a ionizing radiation of \rm $U<10^{-2.5}$. 
Moreover, \citet{matsuoka08} compiled \oi\ and \caii\ line properties of 11 quasars in a broad range of redshift and luminosity ($0.06<z<1.08$ and -$29.8<M_B<-22.1$) in order to analyze the physical conditions that lead to the production of these lines.
They found that the widths of the \oi\  were remarkably similar over more than 3 orders of magnitude in luminosity, suggesting a similar kinematics for location of the emission region. 
They also argue about the dust presence and suggest that the emission region should be located near the dust sublimation point at the outer edge of the BLR.
Thus, the results for \phl\ in Figure~\ref{fig:fwhm}(a) suggests that \feii\ is formed in outer portion of the BLR. 

In contrast, the hydrogen lines are likely to be produced closer to the central source.
Figure~\ref{fig:fwhm}(b) shows the FWHM of \feii\ versus the FWHM of \pab (\pag\ for \phl).
We see that the FWHM of \pab\ is systematically broader than that of \feii. 
For \phl, the FWHM of \hb\ is about $\sim60\%$ larger than that of \feii\ (and the other low-ionization lines analyzed, e.g., \oi\ and \caii).
From the virial theorem, we have that the distance $\rm D \propto 1/\rm FWHM^2$, which translates in a \feii\ emission region 2.5 times more distant than that where the bulk of Hydrogen lines are produced.
\citet{marinello16}	obtained an averaged FWHM for \feii\ that is 3/4 smaller than that of \pab. In particular, for IZw1, the strongest \feii\ emitter of their sample, the results are quite similar to \phl, with the FWHM(\feii)$=1/2~$FWHM(\pab).

Our results are also consistent with the BLR scenario proposed by \citet{dultzin99}. Their BLR model interpret the observed \civ\ as a radial outflow from the central source while low ionization species, such as \feii\ and \caii, arise in the outer portion of the BLR. The intensity and blueshift of \civ\ in this case would depend on the observing angle. In this scenario \civ\ would be produced in a ``cone shaped" inner region. The results for the FWHM(\civ) and strong blueshifted component, narrower \feii, \oi, and \caii\ profiles measured not only in \phl\ but in a larger sample of objects \citep{marinello16} agree with this scenario. 

%In summary, the results above point out that the region emitting \oi, \caii\ and \feii\ in \phl\ is located farther out from the center as compared to that emitting \hb. 
Note that in the above analysis we did not consider any effects between ionization potential of the ion, FWHM and location of the emitting region as is usually considered for \civ\ and Hydrogen lines because all the lines here have very similar ionization potential, $\leq13.6$\,eV.

\subsection{Physical conditions within the BLR}
\label{physblr}

Besides the region where the \feii\ emitting clouds is located relative to the central source, we can also estimate its the physical conditions within the BLR.
The line fluxes from the NIR and UV spectra derived in section~\ref{BLRspec} can be used in combination with CLOUDY simulations to constrain the physical properties of the BLR gas such as the ionization parameter (U) and the gas density (n$_{\rm H}$).

Due to the complexity to model the \feii\ ion, a valid approach is to consider the \oi\ and \caii\ ions as proxies of the former, under the assumption that they all form in clouds that are  co-spatial.
To analyze the role of the excitation mechanisms of \oi\ and the physical conditions of the  region emitting this line, \citet{matsuoka07} published simulations to probe the density and ionization parameter of this region.
A similar approach was followed by \citet{matsuoka08} and more recently by \citet{aldama15}.
Their results are based on the emission line flux ratios  \oi$~\lambda$11287/$\lambda$8446 and the ratio between the Calcium triplet and \oi$~\lambda$8446. The first one is related to the role of the Ly$\beta$ fluorescence in the \oi\ emission while the second one traces the role of collisional excitation. Their results show that the \oi, \caii\ and \feii\ lines arise in a region best characterized with log(n$_{\rm H}$)=11.5~cm$^{-3}$ and log(U)=-2.5. These values are similar to those found in weak to moderate \feii\ emitters.

The first interesting fact about \oi\,$\lambda8446$ and the \caii\ triplet is that the relative intensities of these lines in \phl\ are different from what is usually observed in the literature.
While \oi\,$\lambda8446$ is usually stronger than the \caii\ triplet \citep{riffel06,landt08}, \phl\ shows much stronger \caii\ multiplets. 
For instance, \citet{matsuoka08} analyzing a sample of 11 quasars found a range of line ratios $0.2<$\oi~$\lambda$11287/$\lambda$8446$<0.8$ and $0.3<$\caii/\oi$<0.8$.
\phl\ has similar \oi\ line ratios, \oi$\lambda$11287/$\lambda$8446$=0.6$, but much higher Calcium triplet to \oi\,$\lambda8446$ flux ratio, \caii/\oi$=3.0$.
Indeed, each \caii\ line from the triplet has individual fluxes roughly equal to that of \oi. 

Figure~7 from \citet{matsuoka07} shows isocontour curves in the U,n$_{\rm H}$ plane using the above line ratios.
Based on the \oi\ line ratio measured in \phl, the ionization parameter and density are limited to $-2.4<{\rm log}~U<-4$ and $11.5<$log n$_{\rm H}<14$~cm$^{-3}$, respectively.
In addition, the Calcium to Oxygen line ratio measured for \phl\ implies $-2.0<$U$<-4$ and $11.7<$n$_{\rm H}<14$~cm$^{-3}$.
These values can be further constrained using the models presented by \citet{matsuoka08}. The line ratios found from \phl\ place it in the parameter space consistent to log(U,n$_{\rm H}$)$=$($-3.0,12.5$~cm$^{-3}$).

\citet{negrete12} followed a different approach by considering UV line ratios. 
The set of diagrams presented by them involves flux ratios between emission lines of high- and mid-ionization potential such as Al\,{\sc iii}/Si\,{\sc iii}], Si\,{\sc iv}/Si\,{\sc iii}, C\,{\sc iv}/Al\,{\sc iii}, C\,{\sc iv}/Si\,{\sc iii}], and Si\,{\sc iv}/C\,{\sc iv}.

Using Figure~5 of \citet{negrete12} and the observed UV line flux ratios in \phl\ listed in Table~2, we constrain the physical conditions in the BLR for every ratio employed. In order to achieve convergence, we also followed the approach of \citet{negrete12} (see their Figure~6). 
Given the error in the measurements, we found a convergence point at log(U,n$_{\rm H}$)=($-3.5,13.0$ ~cm$^{-3}$).
These values are slightly different from those obtained for I\,Zw\,1, log(U,n$_{\rm H}$)=($-2.0,12.6$ ~cm$^{-3}$) by \citet{garcia12} and \citet{negrete12}.
Note however that \phl\ spectrum differs significantly from that of I\,Zw\,1 (and other AGN with strong or weaker \feii\ emission).

Are the above conditions derived from the UV lines consistent to those where the \feii\ lines are formed in \phl? In order to answer this question, we employed the predicted \feii\ ratios listed in Table~2 of \citet{garcia12}, based on \citet{sigut03} models. We found that for log(U,n$_{\rm H}$)=($-3.0,12.6$~cm$^{-3}$), the predicted \feii\ line ratio 10502+11127/9200 \AA\ is 0.35. In \phl\, it amounts to 0.45, in very good agreement with the models.  Other combinations of values of U and n$_{\rm H}$ predict values in that ratio that that departs considerably from the observations.  

From the results above we see that a combination of low ionization parameter and high gas density are necessary within the BLR to produce the observed emission lines.
These results are consistent with the parameters derived by \citet{aldama18} for a sample of extreme population of quasars at high redshift.
Their results show that weak \civ\ and C\,{\sc iii}] emission, strong Al\,{\sc iii} and a Si\,{\sc iv}/C\,{\sc iv} line flux ratio $\sim1$ are common properties among quasars of extreme population A, to which \phl\ belongs.
The similar physical properties derived for \phl, which is a local source, with those found in objects at high redshift, suggest that the line ratio diagrams presented by \citet{negrete12} can be used as a selection method to search for extreme \feii\ emitters candidates both in the local universe and at higher redshifts (up to z$\sim$3).

With the launching of the \textit{James Webb Space Telescope} in the next few years, this method can be used as a proxy to identify strong/extreme \feii\ sources at a large stretch of cosmic age.
There has been a lot of attention concerning the sources with R4570>1.0, as these sources (extreme Population A, or xA) seem to radiate at or toward   a limiting Eddington ratio. This property can be exploited, under several assumptions for the use of these sources as cosmological probes \citep{wang13,marziani14}.

%------------------------------------
\subsection{Excitation Mechanisms}

Models employed to study the micro-physics of the \feii\ transitions in AGN incorporate three main excitation mechanisms: collisional excitation, continuum fluorescence via UV resonance lines, and self fluorescence via overlapping \feii\ transitions \citep[see][for a compreensive formulation of the problem]{pradhan11}.
\citet{sigut98} demonstrated that Lymann-$\alpha$ (\lya) fluorescence has an important role in the production of the optical \feii\ lines.
Their results show a significant increase in the strength of the 4570\AA{} bump   when \lya\ fluorescent is fully considered, in comparison when it is weakly (10$\%$) or not included in the calculations.

The key aspect to confirm the fluorescent route for the \feii\ emission is the presence of a bump of emission centered in $\lambda$9200 and the detection of the 1\,$\mu$m \feii\ lines \citep{rudy00, ardila02, marinello16}. 
The role of the different excitation mechanisms and physical conditions necessary to produce the \feii\ emission can be investigated by analyzing the observed emission in different regions of the spectrum. NIR probes \lya\ fluorescence and collisional excitation to energy levels of up to $\sim15$~eV. Optical spectroscopy probes odd parity transitions of \feii\ up $\sim5$~eV while UV transitions, in addition, carries information about the density and ionization parameter.

\begin{figure}
     \includegraphics[width=0.8\columnwidth]{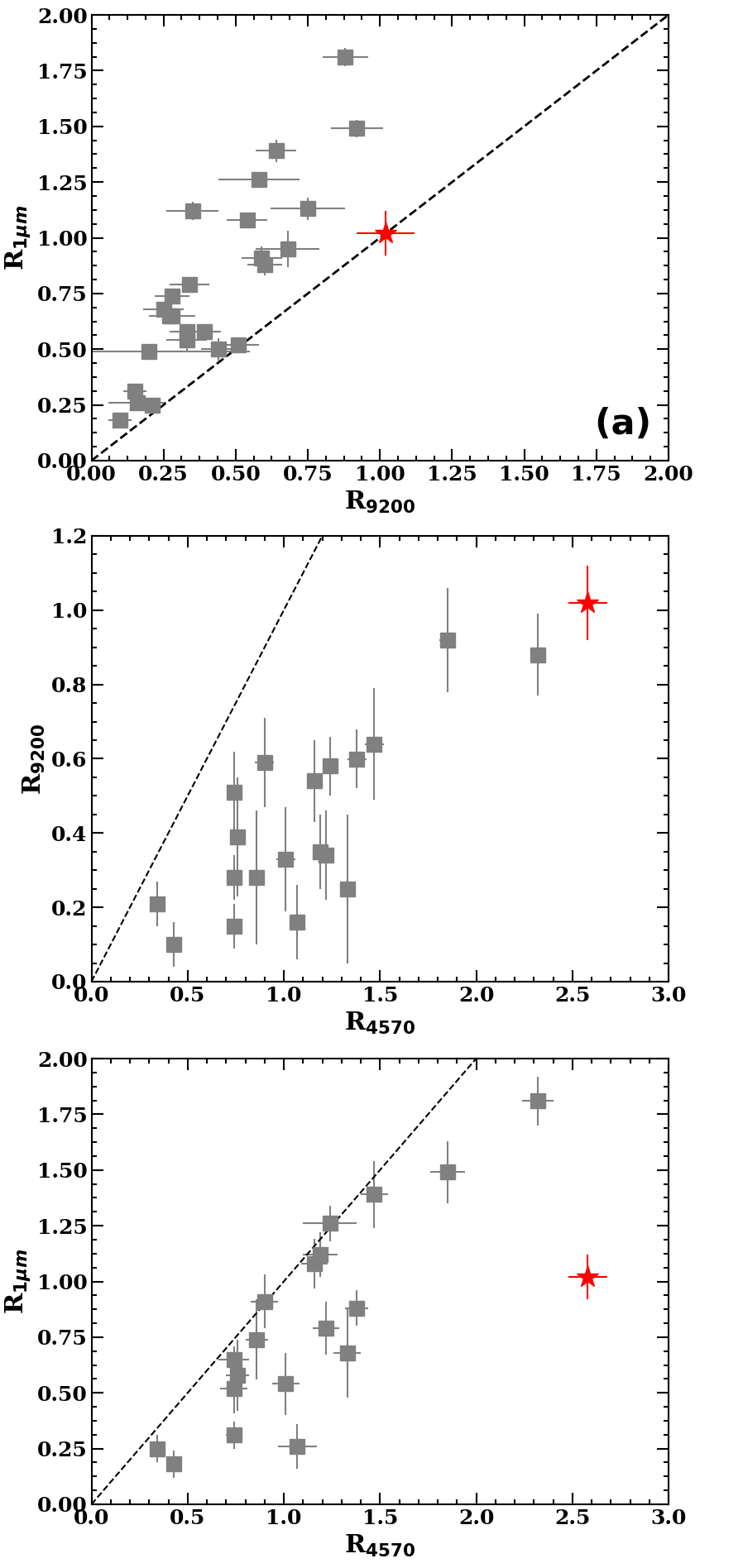}
     \caption{\feii\ line ratios with respect to \hb\ and \pab\ in the optical and NIR, respectively. 
     Panel (a) shows the correlation between the two groups of NIR \feii\ lines important to this work, \rni\ and \ron.
     Panel (b) shows the correlation between \rni\ and the optical \feii\ emission, \rop.
     Panel (c) shows the correlation \ron\ versus \rop. 
     In all panels, the grey squares represent the data from \citet{marinello16}, the red star denotes \phl, and the dashed black line is the equality line.}
     \label{fig:bumps}
\end{figure}

In order to study the \feii\ excitation mechanisms we follow the approach of \citet{marinello16}, which consists of using line flux ratios between \feii\ lines and the closest \ion{H}{i} lines in the same spectrum. The observable quantities are: ($i$) %the flux ratio between the \feii\ bump centered at 4570\AA{} and the broad component of H$\beta$, 
R$_{4570}=\rm F$(\feii\ $\rm 4570)/F(H\beta)$; ($ii$) the flux ratio of the \feii\ bump centered at 9200\AA{} and the broad component of Pa$\gamma$, R$_{9200}=\rm F$(\feii\ $\rm 9200)/F(Pa\gamma)$; and ($iii$) the flux ratio of the \feii\,1$\mu$m lines ($\lambda9998, \lambda10502, \lambda10890, \lambda11127$) and the broad component of Pa$\gamma$, R$_{\rm 1\mu m}=\rm F$(\feii\ $\rm 1\mu m)/F(Pa\gamma)$. Figure~\ref{fig:bumps}, shows the diagrams involving these quantities for \phl\ and other AGN from the literature.

\begin{figure*}
     \includegraphics[width=1.0\textwidth]{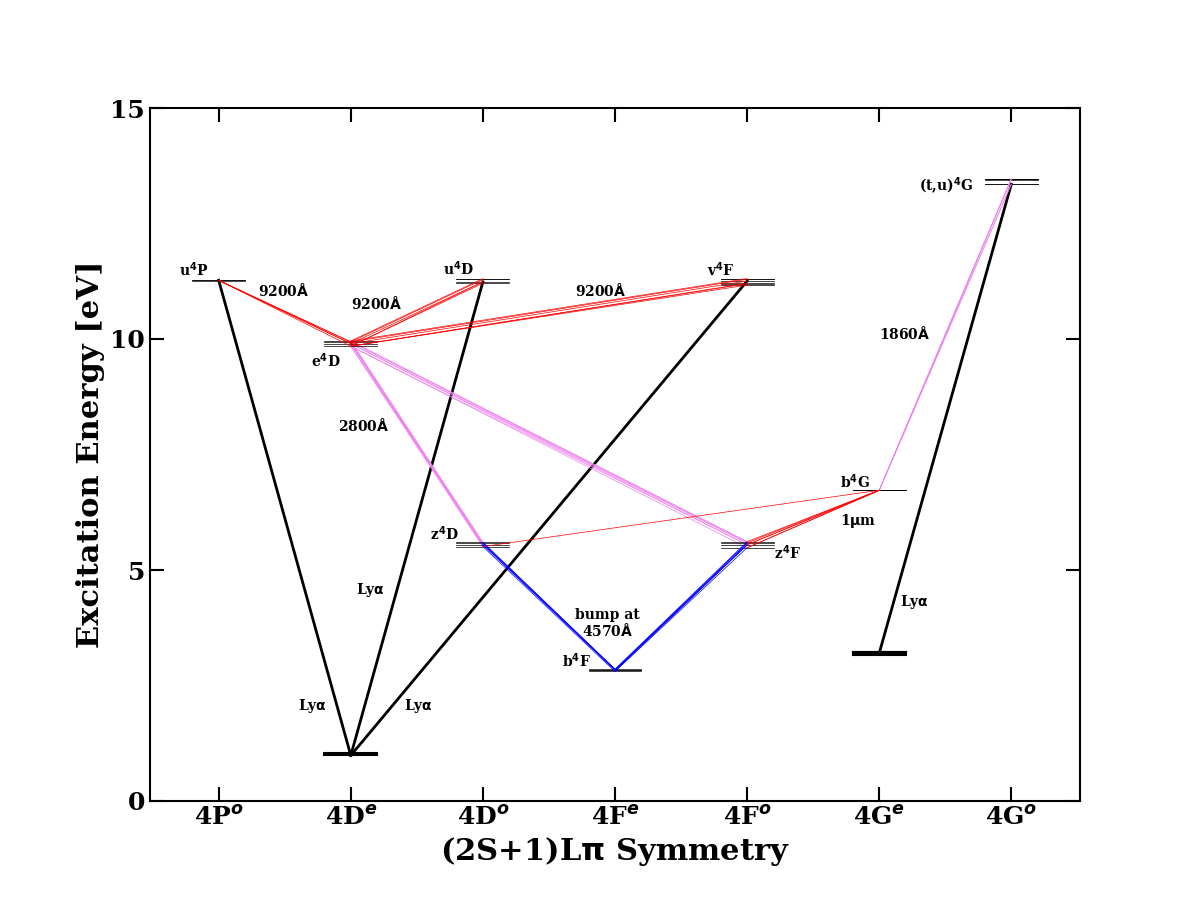}
     \caption{Partial Grotrian diagram for the \feii\ emission. The diagram show the main transition that from primary and/or secondary cascading after a \lya\ photon is absorbed that populate the z$^4$(D,F) levels, from which the optical \feii\ bump (centered at 4570\AA{}) is emitted via z$^4$(D,F)$\rightarrow$b$^4$F$^e$ 
     (blue line). Red lines show the NIR primary and secondary cascade transitions after a \feii\ ion absorbs a \lya\ photon, e.g. the bump at 9200\AA{} and the 1$\mu$m lines. Violet lines represent the UV transitions that populated directly or indirectly the z$^4$(D,F) levels.}
     \label{fig:grotrian}
\end{figure*}

The optical \feii\ emission arises from photons emitted via levels z$^4$(D,F)$\rightarrow$b$^4$F$^e$. Figure~\ref{fig:grotrian} shows a partial Grotrian diagram for the most prominent transitions of \feii\   \citep{sigut03,pradhan11}. In the diagram, red lines represent transitions leading to the bump at 9200\AA{}, violet lines show transitions responsible for the emission lines around 2800\AA{} and 1860\AA{} while blue lines show the transitions that form the bump centered at 4570\AA{}. Table~\ref{tab:transitions} summarizes the main transitions that produce the NIR to UV \feii\ emission. Note that other transitions are still present but they are weaker by a factor of 10 or more and for this reason they are not shown.

\begin{table}
    \centering
    \caption{\feii\ transitions responsible for the bump centered at 4570\AA{}.}
    \label{tab:transitions}
    \begin{tabular}{lccc} % four columns, alignment for each
      \hline
      \hline
      Wavelength         			& Transitions 		&E$_u$ (eV) & E$_l$ (eV) \\  
      
    \hline
    & NIR Transitions&\\
    \hline
    11125.573	&b$^4$G$^{\rm e}_{\rm 5/2}\rightarrow$z$^4$F$^{\rm o}_{\rm 3/2}$ 	&6.727	&5.613 \\
    10862.646	&b$^4$G$^{\rm e}_{\rm 7/2}\rightarrow$z$^4$F$^{\rm o}_{\rm 5/2}$ 	&6.727	&5.587 \\
    10501.521	&b$^4$G$^{\rm e}_{\rm 9/2}\rightarrow$z$^4$F$^{\rm o}_{\rm 7/2}$ 	&6.726	&5.546 \\
    9997.556	&b$^4$G$^{\rm e}_{\rm 11/2}\rightarrow$z$^4$F$^{\rm o}_{\rm 9/2}$ 	&6.721	&5.482 \\
    9377.042	&u$^4$P$^{\rm o}_{\rm 1/2}\rightarrow$d$^4$D$^{\rm e}_{\rm 3/2}$	&11.263	&9.937	\\
    9325.115	&u$^4$P$^{\rm o}_{\rm 3/2}\rightarrow$d$^4$D$^{\rm e}_{\rm 3/2}$	&11.270	&9.937	\\
    9296.851	&u$^4$D$^{\rm o}_{\rm 5/2}\rightarrow$d$^4$D$^{\rm e}_{\rm 5/2}$	&11.238	&9.900	\\
    9210.938	&u$^4$D$^{\rm o}_{\rm 1/2}\rightarrow$d$^4$D$^{\rm e}_{\rm 1/2}$	&11.307	&9.957	\\
    9203.122	&v$^4$F$^{\rm o}_{\rm 3/2}\rightarrow$d$^4$D$^{\rm e}_{\rm 1/2}$	&11.308	&9.957	\\
    9196.897	&u$^4$D$^{\rm o}_{\rm 3/2}\rightarrow$d$^4$D$^{\rm e}_{\rm 3/2}$	&11.288	&9.937	\\
    9178.008	&v$^4$F$^{\rm o}_{\rm 5/2}\rightarrow$d$^4$D$^{\rm e}_{\rm 3/2}$ 	&11.291	&9.937  \\
    9175.869	&v$^4$F$^{\rm o}_{\rm 7/2}\rightarrow$d$^4$D$^{\rm e}_{\rm 5/2}$ 	&11.255	&9.900  \\
    9132.362	&v$^4$F$^{\rm o}_{\rm 9/2}\rightarrow$d$^4$D$^{\rm e}_{\rm 7/2}$ 	&11.206	&9.845  \\
    9122.942	&u$^4$D$^{\rm o}_{\rm 7/2}\rightarrow$d$^4$D$^{\rm e}_{\rm 7/2}$ 	&11.208	&9.845  \\
    9077.400	&u$^4$P$^{\rm o}_{\rm 3/2}\rightarrow$d$^4$D$^{\rm e}_{\rm 5/2}$ 	&11.270	&9.900  \\
    9075.501	&v$^4$P$^{\rm o}_{\rm 5/2}\rightarrow$d$^4$D$^{\rm e}_{\rm 5/2}$ 	&11.270	&9.900  \\
    8926.638	&u$^4$D$^{\rm o}_{\rm 5/2}\rightarrow$d$^4$D$^{\rm e}_{\rm 7/2}$ 	&11.238	&9.845  \\

      \hline
     & Optical Transitions&\\
      \hline
    4629.337	&z$^4$F$^{\rm o}_{\rm 9/2}\rightarrow$b$^4$F$^{\rm e}_{\rm 9/2}$ 	&5.482	&2.805 \\    
    4583.822	&z$^4$D$^{\rm o}_{\rm 7/2}\rightarrow$b$^4$F$^{\rm e}_{\rm 9/2}$ 	&5.508	&2.805 \\
    4582.831	&z$^4$F$^{\rm o}_{\rm 7/2}\rightarrow$b$^4$F$^{\rm e}_{\rm 5/2}$ 	&5.546	&2.843 \\
    4576.339	&z$^4$D$^{\rm o}_{\rm 5/2}\rightarrow$b$^4$F$^{\rm e}_{\rm 5/2}$ 	&5.550	&2.843 \\
    4555.878	&z$^4$F$^{\rm o}_{\rm 7/2}\rightarrow$b$^4$F$^{\rm e}_{\rm 7/2}$ 	&5.546	&2.827 \\
    4549.462	&z$^4$D$^{\rm o}_{\rm 5/2}\rightarrow$b$^4$F$^{\rm e}_{\rm 7/2}$ 	&5.550	&2.827 \\
    4541.524	&z$^4$D$^{\rm o}_{\rm 3/2}\rightarrow$b$^4$F$^{\rm e}_{\rm 3/2}$ 	&5.582	&2.854 \\
    4522.632	&z$^4$D$^{\rm o}_{\rm 3/2}\rightarrow$b$^4$F$^{\rm e}_{\rm 5/2}$ 	&5.582	&2.843 \\
    4520.207	&z$^4$F$^{\rm o}_{\rm 7/2}\rightarrow$b$^4$F$^{\rm e}_{\rm 9/2}$ 	&5.546	&2.805 \\
    4515.345	&z$^4$F$^{\rm o}_{\rm 5/2}\rightarrow$b$^4$F$^{\rm e}_{\rm 5/2}$ 	&5.587	&2.843 \\
    4508.284	&z$^4$D$^{\rm o}_{\rm 1/2}\rightarrow$b$^4$F$^{\rm e}_{\rm 3/2}$ 	&5.603	&2.854 \\
    4491.414	&z$^4$F$^{\rm o}_{\rm 3/2}\rightarrow$b$^4$F$^{\rm e}_{\rm 3/2}$ 	&5.613	&2.854 \\
    4489.179	&z$^4$F$^{\rm o}_{\rm 5/2}\rightarrow$b$^4$F$^{\rm e}_{\rm 7/2}$ 	&5.587	&2.827 \\

    \hline
    & UV Transitions&\\
    \hline    
    2884.764	&d$^4$D$^{\rm e}_{\rm 7/2}\rightarrow$z$^4$D$^{\rm o}_{\rm 5/2}$ 	&9.845	&5.550 \\
    2882.190	&d$^4$D$^{\rm e}_{\rm 7/2}\rightarrow$z$^4$F$^{\rm o}_{\rm 7/2}$ 	&9.845	&5.546 \\
    2872.258	&d$^4$D$^{\rm e}_{\rm 5/2}\rightarrow$z$^4$F$^{\rm o}_{\rm 5/2}$ 	&9.900	&5.587 \\
    2869.317	&d$^4$D$^{\rm e}_{\rm 5/2}\rightarrow$z$^4$D$^{\rm o}_{\rm 3/2}$ 	&9.900	&5.582 \\

    2858.629	&d$^4$D$^{\rm e}_{\rm 3/2}\rightarrow$z$^4$D$^{\rm o}_{\rm 1/2}$ 	&9.937	&5.603 \\
    2856.908	&d$^4$D$^{\rm e}_{\rm 7/2}\rightarrow$z$^4$D$^{\rm o}_{\rm 7/2}$ 	&9.845	&5.508 \\
    2856.454	&d$^4$D$^{\rm e}_{\rm 3/2}\rightarrow$z$^4$F$^{\rm o}_{\rm 3/2}$ 	&9.937	&5.613 \\

    2851.721	&d$^4$D$^{\rm e}_{\rm 1/2}\rightarrow$z$^4$F$^{\rm o}_{\rm 3/2}$ 	&9.957	&5.613 \\
    2848.315	&d$^4$D$^{\rm e}_{\rm 3/2}\rightarrow$z$^4$F$^{\rm o}_{\rm 5/2}$ 	&9.937	&5.587 \\
    2848.110	&d$^4$D$^{\rm e}_{\rm 5/2}\rightarrow$z$^4$D$^{\rm o}_{\rm 5/2}$ 	&9.900	&5.550 \\
    2845.601	&d$^4$D$^{\rm e}_{\rm 5/2}\rightarrow$z$^4$F$^{\rm o}_{\rm 7/2}$ 	&9.900	&5.546 \\

    2839.507	&d$^4$D$^{\rm e}_{\rm 9/2}\rightarrow$z$^4$F$^{\rm e}_{\rm 9/2}$ 	&9.845	&5.482 \\
    2831.881	&d$^4$D$^{\rm e}_{\rm 1/2}\rightarrow$z$^4$D$^{\rm e}_{\rm 3/2}$ 	&9.957	&5.582 \\
    2824.566	&d$^4$D$^{\rm e}_{\rm 3/2}\rightarrow$z$^4$D$^{\rm e}_{\rm 5/2}$ 	&9.937	&5.550 \\
    2820.954	&d$^4$D$^{\rm e}_{\rm 5/2}\rightarrow$z$^4$D$^{\rm e}_{\rm 7/2}$ 	&9.900	&5.508 \\

    1877.716	&u$^4$G$^{\rm o}_{\rm 7/2}\rightarrow$b$^4$G$^{\rm e}_{\rm 9/2}$ 	&13.348	&6.726 \\
    1872.973	&u$^4$G$^{\rm o}_{\rm 9/2}\rightarrow$b$^4$G$^{\rm e}_{\rm 7/2}$ 	&13.350	&6.727 \\
    1872.638	&u$^4$G$^{\rm o}_{\rm 9/2}\rightarrow$b$^4$G$^{\rm e}_{\rm 9/2}$ 	&13.350	&6.726 \\
    1869.553	&u$^4$G$^{\rm o}_{\rm 11/2}\rightarrow$b$^4$G$^{\rm e}_{\rm 11/2}$ 	&13.356	&6.721 \\
    1843.260	&t$^4$G$^{\rm o}_{\rm 11/2}\rightarrow$b$^4$G$^{\rm e}_{\rm 11/2}$ 	&13.450	&6.721 \\
    1841.710	&t$^4$G$^{\rm o}_{\rm 9/2}\rightarrow$b$^4$G$^{\rm e}_{\rm 9/2}$ 	&13.461	&6.726 \\
    1839.999	&t$^4$G$^{\rm o}_{\rm 7/2}\rightarrow$b$^4$G$^{\rm e}_{\rm 7/2}$ 	&13.468	&6.727 \\
    1839.802	&t$^4$G$^{\rm o}_{\rm 5/2}\rightarrow$b$^4$G$^{\rm e}_{\rm 5/2}$ 	&13.468	&6.727 \\
    \hline
    \hline
    \end{tabular}
\end{table}

The 9200\AA{} bump is a blend of lines produced by primary cascading from the energy levels around 13~eV, excited by the capture of a \lya\ photon by an \feii\ ion, via u$^4$(P,D)$\rightarrow$e$^4$D and v$^4$F$\rightarrow$e$^4$D.
Secondary cascading occurs from the levels e$^4$D, at about 10~eV, to the upper levels from which the optical emission lines are produced, after the emission of UV photons at $\sim$2800\AA{}, via e$^4$D$\rightarrow$z$^4$(D,F). 
Since the energy to excite the u$^4$(P,D) and v$^4$F  levels are too high to be excited by mechanisms other than \lya\ fluorescence, the detection of the 9200\AA{} bump probes unambiguously the presence of that mechanism.

Figure~\ref{fig:bumps}(a) shows \rni\ versus \ron. The grey dots from \citet{marinello16} suggest a correlation between these two ratios.
The inclusion of \phl\ in the plot introduces some scatter to the high end of the correlation.
The 1$\mu$m lines result from secondary cascading of the \lya\ fluorescence process, e.g., after a \feii\ ion captures a \lya\ photon, electrons are excited to levels at $\sim$15eV, (t,u)$^4$G, which cascade down via (t,u)$^4$G$\rightarrow$b$^4$G emitting photons at around 1860\AA{}.
In the secondary cascading, the 1$\mu$m lines are emitted via b$^4$G$\rightarrow$z$^4$(D,F), populating the energy levels from which the bulk of the optical \feii\ is produced.
Note that the b$^4$G level needs only $\sim$6~eV to be excited, thereby collisional excitation is also a possible mechanism to pump this level.
The strong correlation observed in Figure~\ref{fig:bumps}(a) shows the importance of the \lya\ fluorescence in the production of the 1$\mu$m \feii\ lines.
%An interest aspect of the \phl\ at the plot is that it falls over the unitary line.

\citet{garcia12} used a grid of \feii\ models in the NIR in order to create an \feii\ template. 
They found that models with log(U,$n_{\rm H}$)=(-2.0,12.6~cm$^{-3}$) best reproduce the \feii\ emission in I\,Zw\,1.
In their models, the strength of the 1$\mu$m lines is reduced relative to that of the 9200\AA{} bump for models with the parameters we derived in this work, e.g., log(U,$n_{\rm H}$)=(-3.0,13.0~cm$^{-3}$).
This reduction is indeed observed in \phl, producing its departure from the main trend observed in the Figure~\ref{fig:bumps}(a).

Figure~\ref{fig:bumps}(b) shows \rop\ versus \rni.
This plot strengthens the importance of the \lya\ fluorescence in the production of the optical \feii\ emission.
Since the 9200\AA{} bump is produced exclusively by that mechanism, the plot gives us a rough estimate of the percentage of its contribution to the optical emission. 

\citet{marinello16} found a strong correlation between the above two ratios. However, their sample did not include extreme strong \feii\ emitters (\rop$>2.5$).
We see that the correlation still holds after the inclusion of \phl\ in the diagram.
The physical conditions derived in the previous section, log(U,$n_{\rm H}$)=(-3.0,13.0~cm$^{-3}$), enhance the \lya\ fluorescence. 
Very likely, the \rni\ measured in \phl\ is expected to be close to the maximum possible for the given Pa$\beta$ flux.
In \phl\ the contribution of \lya\ fluorescence to the \feii\ emission is not only important, as in AGN with weak \feii, but also crucial to enhance the \feii\ bump at 4570\AA{}.
Since the 4570\AA{} bump can be pumped by other excitation mechanisms (i.e. collisional excitation) we would expect, in theory, that it keeps increasing while \rni\ should remain roughly constant, with values around one, at the high end of the correlation.
Note that the three strong \feii\ emitters (\rop$>2.0$) displayed in the plot have similar \rni\ as \rop\ grows.
If we assume that the optimal physical conditions for \lya\ fluorescence is met, a maximum value for the \rni\ should be reached while the \rop\ keeps growing towards \rop$\sim3$.
This behaviour is expected from the perspective of photon balance and the intrinsic cascading process of \lya\ fluorescence, because every photon from the 9200\AA{} blend produces a photon in the 4570\AA{} bump under the assumption of zero extinction.
Note that a photon emitted by the 9200\AA{} bump has an energy equals to $E_{9200\AA}=2.16\times10^{-12}$\,erg while a photon from 4570\AA{} has $E_{4570\AA}=4.35\times10^{-12}$\,erg, roughly the double of the energy.
This means that it is necessary twice as much energy to increase \rop\ compared with that of \rni.
Therefore, a flattening in the high end of the correlation is expected.
%A larger sample of extreme \feii\ emitters is necessary to confirm this theoretical argument.

Figure~\ref{fig:bumps}(c) shows \rop\ versus \ron.
The importance of this plot is on the fact that the levels z$^4$(D,F), from which the optical \feii\ emission is produced, are populated after the 1$\mu$m lines are emitted via b$^4$G$\rightarrow$z$^4$(D,F). 
The energy necessary to excite b$^4$G is $\sim6$eV and can be reached by two main excitation mechanisms: collisional excitation and \lya\ fluorescence \citep{marinello16}.

Figure~\ref{fig:bumps}(a) confirms the trend between \rni\ and \ron, with some scatter for strong \feii\ emitters, suggesting \lya\ fluorescence as an important mechanism for the 1$\mu$m lines.
Moreover, for \phl\ the values of \rni\ and \ron\ are similar, which suggests that \lya\ fluorescence must have an even more importance in the formation of these lines.
If we assume that both transitions have similar probability, then the contribution of the \lya\ fluorescence in the production of the 1$\mu$m lines should also be optimal for these lines for the physical conditions within the BLR.
Note that as in Figure~\ref{fig:bumps}(b) we cannot expect the correlation to grow to extreme values of \ron.
The discussion above made for Figure~\ref{fig:bumps}(b) can also be applied for this case since the 1$\mu$m photons have similar energy ($E_{1\mu m}=1.99\times10^{-12}$erg) that the ones from the 9200\AA{} bump.

The results gathered from Figure~\ref{fig:bumps} suggest that the main excitation mechanism pumping the \feii\ emission in \phl\ 
%(and possible all other extreme \feii\ emitters) 
is \lya\ fluorescence. In contrast, weaker \feii\ emitters have their emission driven by collisional excitation, as found by \citet{marinello16}.
Note that other parameters may also contribute in the enhancement of the \feii\ emission in AGN.
For instance, \citet{sigut03} showed that the presence of micro-turbulence ($\eta_t$) in the BLR clouds can also increase the efficiency of the \feii\ production (their models uses $\eta_t=10~$\,km\,s$^{-1}$). \citet{garcia12} showed that for that $\eta_t$ and log(U,$n_{\rm H}$)=(-2.0,12.6~cm$^{-3}$) the \citet{sigut03} models reproduce well the \feii\ emission even in I\,Zw\,1, a strong \feii\ emitter.
\citet{bruhweiler08} found that doubling the turbulence, $\eta_t=20$\,km\,s$^{-1}$, in combination with a larger ionization parameter and a slighter lower density, the predicted \feii\ emission also reproduces well the observed I\,Zw\,1 spectrum in optical/UV regions.
Thus, increasing the turbulence velocity increases also the probability of a \lya\ photon to be captured by an \feii\ ion, thereby increasing the relevance of the \lya\ fluorescence in the observed \feii\ emission. This could be the case for \phl. That is, in that source we are  probing a higher turbulence velocity for the gas in the BLR.

Another alternative for the pumping of the \feii\ emission in extreme \feii\ emitters is a higher chemical abundance in the BLR.
Based on simulations, \citet{panda18} showed that increasing the BLR metallicity leads to a significant increase in \rop.
They showed that a $Z=3$\,$Z_{\odot}$ produces nearly twice the \feii\ emission than at solar metallicity. Indeed, an increase of \rop\ by a factor of 3 can be obtained when $Z=10$\,$Z_{\odot}$.

\citet{negrete12} analyzed the physical condition of the BLR in two strong \feii\ emitters, I\,Zw\,1 and SDSS\,J12014+0116. They found that an abundance five times solar in Aluminum and Silicon was associated to a higher ionization parameter and a lower density, log(U,$n_{\rm H}$)=(-2.75,12.3 cm$^{-3}$). These values, consistent with \citet{garcia12} and \citet{sigut03} models, would reproduced the observed UV line ratios in the spectra.
One possible route to increase the metallicity in the BLR would be through supernova activity in the circunuclear region, whose ejecta could enrich the AGN surrounding media.
The over-solar matter then would be transported towards the BLR via inflows.

Observations give support to this scenario. \citet{watabe08} had already found that the nuclear starburst luminosity in nearby AGN was dependent on the AGN Eddington ratio while \citet{hennig18} found in the local NLS1 galaxy, Mrk\,42, a starburst nuclear ring.  
Another possibility is that the strong outflows detected in \civ\ and Si\,{\sc iv} by means of blueshifted components, swept away heavy elements, thereby increasing the metallicity. Atoms such as Carbon and Silicon, are efficiently accelerated by resonance line scattering (i.e., in a ``line driven wind" scenario). Therefore, the outflowing gas may appear enriched with respect to the accretion disk gas \citep{baskin12}.
Since the models of \citet{sigut03} and the template of \citet{garcia12} were constructed for Solar abundances, the higher abundance scenario cannot be currently tested.
However, the models presented in \citet{garcia12} show a clear increase in the 9200\AA{} bump for the physical parameters obtained in this work and are also consistent with the line ratios in the observed spectrum of \phl. These results support that a lower ionization parameter and a higher density are the most likely  cause of the increase efficiency of the \lya\ fluorescence in \phl.
Simulations with a more focused grid of parameters, high spatial resolution observations, and a bigger sample of extreme \feii\ emitters are necessary to draw more robust conclusions about these hypothesis.

%%%%%%%%%%%%%%%%%%%%%%%%%%%%%%%%%%%%%%%%%%%%%%%%%%
%%%%%%%%%%%%%% END OF DISCUSSION %%%%%%%%%%%%%%%%%
%%%%%%%%%%%%%%%%%%%%%%%%%%%%%%%%%%%%%%%%%%%%%%%%%%

%-------------------------------------------------

\section{Conclusions}

In this work we present a panchromatic (UV, optical and NIR) analysis of the extreme strong \feii\ emitter \phl. 
We combine optical and NIR spectroscopic observations to estimate several physical properties of this outstanding source.
Using template modeling of the \feii\ spectrum, we were able to estimate accurately the intensity of that emission.
Emission lines were fitted using a Lorentzian model in order to obtain their fluxes and line widths.
The main results obtained from these measurements are summarized below:
\begin{itemize}
    \item We re-estimated the \rop\ ratio in \phl\ using a more robust approach. Our results show a \rop =2.58, significantly smaller than the values reported previously, of up to \rop=6.2. This new value, however, still places \phl\    among the strongest \feii\ emitters ever observed.
    \item The FWHM obtained for \feii, \oi, and \caii\ suggest that the clouds emitting these ions are co-spatial, since they share roughly the same line width and shape. Compared with \hb, the FWHM of these lines are significantly smaller, suggesting that the clouds emitting low ionization ions are in the outer portion of the BLR while \hb\ is emitted in the mid-portion of that region.
    \item Emission line flux ratios in the  and UV in \phl\ suggest high densities ($n_{\rm H}\sim 10^{13.0}$~cm$^{-3}$) and a low ionization parameter (log $U \sim -3.5$) for the \feii\ emitting gas. This result is consistent with the parameters employed to create the NIR \feii\ template in \citet{garcia12}.
    \item We found an excess of \feii\ emission in the 9200\AA{} bump. Since this emission can only be produced by ly$\alpha$-fluorescence, we suggest that this excitation mechanism is pumping up the bulk of the \feii\ emission in \phl. This result contrasts to what is found in weaker \feii\ emitters, where collisional excitation seems to be the main driver of that emission.
    \item The re-estimated value of \rop\ places \phl\ among the extreme quasars in the main quasar sequence defined by the optical plane of the Eigenvector 1. In this plane, \phl\ is classified as {\bf extreme} 'Population A' AGN due its high \rop\ and low FWHM$_{\rm H\beta}$. 
\end{itemize}

Our work points out towards intrinsic differences in the physical conditions of  the BLR in this extreme \feii\ emitter even though the gas distribution follows the same trend as in normal \feii\ emitters. A follow-up investigation of this issue using a larger sample of '\phl-like' sources is necessary to fully support our conclusions on the physical conditions driving the \feii\ emission along the E1 plane.

\section*{Acknowledgements}
A.R.-A. acknowledges partial support from CNPq Fellowship (311935/2015-0 \& 203746/2017-1).
M.M. acknowledges partial support from PCI Fellowship
PM acknowledges funding from the INAF PRIN-SKA 2017 program 1.05.01.88.04. 

%%%%%%%%%%%%%%%%%%%%%%%%%%%%%%%%%%%%%%%%%%%%%%%%%%
%%%%%%%%%%%%%%%%% APPENDICES %%%%%%%%%%%%%%%%%%%%%
%%%%%%%%%%%%%%%%%%%%%%%%%%%%%%%%%%%%%%%%%%%%%%%%%%
%\appendix

% If you want to present additional material which would interrupt the flow of the main paper,
% it can be placed in an Appendix which appears after the list of references.

%%%%%%%%%%%%%%%%%%%%%%%%%%%%%%%%%%%%%%%%%%%%%%%%%%

% Don't change these lines
\bsp	% typesetting comment
\label{lastpage}
\end{document}